\documentclass[groupedaddress,superscriptaddress,aps,prd,preprintnumbers,onecolumn,10pt,fleqn,nofootinbib,nobibnote,eqsecnum,floatfix,showpacs,preprintnumbers]{revtex4-1}

\usepackage{graphicx}
\usepackage{bm,amsmath,amssymb}
\usepackage[mathscr]{eucal}
\usepackage{color}
\usepackage{afterpage}

\long\def\comment#1{ }
\newcommand{\eqn}[1]{Eq.~\eqref{#1}}
\newcommand{\beq}{\begin{equation}}
\newcommand{\eeq}{\end{equation}}
\newcommand{\nn}{\nonumber\\}
\newcommand{\dif}{{\rm d}}
\newcommand{\rmd}{{\rm d}}
\newcommand{\rme}{{\rm e}}
\newcommand{\rmi}{{\rm i}}
\newcommand{\rmP}{{\rm P}}

\newcommand{\rmtr}{{\rm tr}}

\newcommand{\del}{\partial}

\newcommand{\order}[1]{\mcal{O}{(#1)}}
\newcommand{\mcal}{\mathcal}

\newcommand{\bk}{\bm{k}}
\newcommand{\bq}{\bm{q}}

\newcommand{\bel}{\bm{\ell}}
\newcommand{\bx}{\bm{x}}
\newcommand{\by}{\bm{y}}

\newcommand{\bz}{\bm{z}}

\newcommand{\br}{\bm{r}}

\newcommand{\abar}{\bar{\alpha}_s}

\newcommand{\ssE}{{\rm \scriptscriptstyle E}}

\begin{document}
\preprint{TUM-EFT 106/17}

\title{On the use of a running coupling in the NLO calculation of forward hadron production}

\author{B.~Duclou\'{e}}
\email{bertrand.ducloue@ipht.fr}
\affiliation{Institut de Physique Th\'{e}orique, Universit\'{e} Paris-Saclay, CNRS, CEA, F-91191 Gif-sur-Yvette, France}
\affiliation{Department of Physics, 40014 University of Jyv\"{a}skyl\"{a}, Finland\\
and Helsinki Institute of Physics, 00014 University of Helsinki, Finland}

\author{E.~Iancu}
\email{edmond.iancu@ipht.fr}
\affiliation{Institut de Physique Th\'{e}orique, Universit\'{e} Paris-Saclay, CNRS, CEA, F-91191 Gif-sur-Yvette, France}

\author{T.~Lappi}
\email{tuomas.v.v.lappi@jyu.fi}
\affiliation{Department of Physics, 40014 University of Jyv\"{a}skyl\"{a}, Finland\\
and Helsinki Institute of Physics, 00014 University of Helsinki, Finland}

\author{A.H.~Mueller}
\email{amh@phys.columbia.edu}
\affiliation{Department of Physics, Columbia University, New York, NY 10027, USA}

\author{G.~Soyez}
\email{gregory.soyez@ipht.fr}
\affiliation{Institut de Physique Th\'{e}orique de Saclay,
F-91191 Gif-sur-Yvette, France}

\author{D.N.~Triantafyllopoulos}
\email{trianta@ectstar.eu}
\affiliation{European Centre for Theoretical Studies in Nuclear Physics and Related Areas (ECT*)\\ and Fondazione Bruno Kessler, Strada delle Tabarelle 286, I-38123 Villazzano (TN), Italy}

\author{Y.~Zhu}
\email{yanzhu.zhu@tum.de}
\affiliation{Department of Physics, 40014 University of Jyv\"{a}skyl\"{a}, Finland\\
and Helsinki Institute of Physics, 00014 University of Helsinki, Finland}
\affiliation{Physik-Department, Technische Universit{\" a}t M{\" u}nchen, D-85748 Garching, Germany}

\date{\today}

\begin{abstract}

We address and solve a puzzle raised by a recent calculation \cite{Ducloue:2017mpb} of the cross-section for particle production in proton-nucleus collisions to next-to-leading order: the numerical results show an unreasonably large dependence upon the choice of a prescription for the QCD running coupling, which spoils the predictive power of the calculation. Specifically, the results obtained with a prescription formulated in the transverse coordinate space differ by one to two orders of magnitude from those obtained with a prescription in momentum space. We show that this discrepancy is an artefact of the interplay between the asymptotic freedom of QCD and the Fourier transform from coordinate space to momentum space. When used in coordinate space, the running coupling can act as a fictitious potential which mimics hard scattering and thus introduces a spurious contribution to the cross-section. We identify a new coordinate-space prescription which avoids this problem and leads to results consistent with those obtained with the momentum-space prescription.
\end{abstract}

\pacs{
12.38.Bx, 
12.38.Cy, 
12.39.St, 
25.75.-q 
}
\maketitle


\section{\label{sec:intro} Introduction}
Particle production in proton-proton or proton-nucleus collisions at forward rapidities and semi-hard transverse momenta of a few GeV represents an important source of information about the small-$x$ part of the nuclear wavefunction, where gluon occupation numbers are high and non-linear effects like gluon saturation and multiple scattering are expected to be important.  To optimise our extraction of this physical information from the experimental data at RHIC and the LHC, reliable theoretical predictions are of utmost importance. For such semi-hard momenta, the QCD coupling is only moderately weak, hence the perturbative calculations of the relevant cross-sections must be pushed up to next-to-leading order (NLO) accuracy at least.

In the recent years, the Color Glass Condensate (CGC) effective theory \cite{Gelis:2010nm} (the natural framework for such calculations within perturbative QCD) has indeed been promoted to NLO accuracy.  This includes both the equations describing the high-energy evolution of the scattering amplitude in the presence of non-linear phenomena --- the B-JIMWLK\footnote{This acronym stays for Balitsky, Jalilian-Marian, Iancu, McLerran, Weigert, Leonidov and Kovner.} hierarchy of equations \cite{Balitsky:1995ub,JalilianMarian:1997jx,JalilianMarian:1997gr,Kovner:2000pt,Iancu:2000hn,Iancu:2001ad,Ferreiro:2001qy} and its mean field approximation known as the Balitsky-Kovchegov (BK) equation  \cite{Balitsky:1995ub,Kovchegov:1999yj} --- and the impact factors describing the coupling between a dilute projectile and the dense gluon distribution from the nuclear target. However, the first NLO calculations met with serious difficulties (instability of the NLO BK equation, negative cross-section for particle production, huge scheme-dependence in the choice of a prescription for the running of the coupling), that were progressively understood and overcome. 

Specifically, the NLO version of the BK equation \cite{Balitsky:2008zza,Balitsky:2013fea,Kovner:2013ona} turned out to be unstable \cite{Avsar:2011ds,Lappi:2015fma}, due to the presence of large and negative NLO corrections enhanced by double collinear logarithms. Two independent solutions have been proposed to solve this problem --- the enforcement of a kinematical constraint\footnote{See also Ref.~\cite{Hatta:2016ujq} for a generalization of the  kinematical constraint to the Langevin formulation of the JIMWLK equation.} \cite{Beuf:2014uia} and a collinearly-improved version of the BK equation  \cite{Iancu:2015vea,Iancu:2015joa} ---,  which refer to two different methods for resumming the double collinear logarithms to all orders.  After restoring the full NLO accuracy (by adding the remaining NLO corrections, in particular, those expressing the running of the coupling) on top of the collinear improvement \cite{Iancu:2015vea,Iancu:2015joa}, the numerical solutions to the BK equation were found to be stable and physically meaningful \cite{Lappi:2016fmu}.

Furthermore, the first NLO calculations of single-inclusive particle production in proton-nucleus collisions at forward rapidities \cite{Chirilli:2011km,Chirilli:2012jd,Stasto:2013cha,Stasto:2014sea,Xiao:2014uba,Watanabe:2015tja,Ducloue:2016shw} led to a cross-section which suddenly drops and becomes negative when increasing the transverse momentum of the produced hadron (but still within the semi-hard region where the formalism is supposed to apply).
These NLO calculations used the so-called ``hybrid factorization'' \cite{Dumitru:2005gt} together with the NLO correction to the impact factor originally computed by Chirilli, Xiao, and Yuan  \cite{Chirilli:2011km,Chirilli:2012jd} (see also \cite{Altinoluk:2014eka} for an alternative calculation). The word ``hybrid'' refers to the fact that the projectile proton and the nuclear target are treated on a different footing. The proton, which is dilute, is treated within collinear factorization, that is, by using the standard, ``integrated'', parton distributions which also appear in the study of deep inelastic scattering. The dense nucleus, on the other hand, is treated within the spirit of $k_T$--factorization, that is, in terms of  ``unintegrated'' gluon distributions which describe the distribution of the (soft) gluons in both longitudinal and transverse momenta, including its non-linear evolution with increasing energy. The ``impact factor'' encodes the coupling between a collinear parton from the incoming proton and the soft gluons in the nucleus. Within $k_T$--factorization, it is assumed that this impact factor can be computed at a fixed rapidity separation between the projectile and the target, and hence it can be separated from the high-energy evolution. This is however just an approximation: starting with NLO, the calculation includes gluon fluctuations which span the whole rapidity phase-space and thus probe the target gluon distribution within an extended range in rapidity.

As explained in Ref.~\cite{Iancu:2016vyg}, the problem with the negativity of the NLO cross-section mainly  comes from enforcing such a local (in rapidity) separation between the evolution and the NLO impact factor (the ``rapidity subtraction'').  This difficulty is therefore expected to be generic. And indeed, as recently shown in Ref.~\cite{Ducloue:2017ftk}, a similar problem appears in the context of  deep inelastic scattering (DIS), when using the ``dipole factorization'' (a version of $k_T$-factorization appropriate for DIS at high energy) together with  the NLO impact factor from Refs.~\cite{Beuf:2016wdz,Beuf:2017bpd}.

An additional source of difficulty, that will be our main focus in this paper, is the mismatch between the running coupling prescriptions used in the calculation of the NLO impact factor and in the high-energy evolution (the BK equation), respectively. On one hand, the BK equation is most naturally formulated and solved in the transverse {\it coordinate} space. Indeed, the coordinate representation allows for simple implementations of the eikonal approximation and of the unitarity constraint on the dipole scattering amplitude. On the other hand, the cross-section, hence the NLO impact factor, must be ultimately expressed in terms of the transverse {\it momentum} of the produced particle, which is a measurable quantity.\footnote{The transverse momentum representation of  the impact factor  is also convenient in view of the subtraction of the collinear divergences from the one-loop calculation \cite{Chirilli:2012jd}; see the discussion in Sect.~\ref{sec:cf} below.}  This makes it natural to use a coordinate-space scale (a dipole size) for the running coupling in the solution to the BK equation, but a momentum-space scale (a transverse momentum) in the one-loop calculation of the NLO impact factor. (The precise prescriptions that we shall use in practice will be specified later on.) Albeit not a fundamental problem in itself --- the use of various  running coupling (RC) prescriptions in different sectors of the calculation is consistent with the overall NLO accuracy and should be viewed as a part of the {\it scheme-dependence} inherent in the formalism ---, this mismatch amplifies the negativity problem of the NLO cross-section \cite{Iancu:2016vyg}. In fact, it may introduce such a problem by itself, even when the separation between the leading-order result and the NLO corrections is made by properly keeping the non-locality of the one-loop calculation in rapidity. 

To avoid such problems altogether, Ref.~\cite{Iancu:2016vyg} proposed an ``unsubtracted'' factorization scheme which avoids the subtraction of the leading-order evolution from the full NLO result. This factorization preserves the actual skeleton structure of the original Feynman graphs: the ``impact factor'' is now given by a full one-loop graph in which the soft (small-$x$) part of the gluon fluctuation contributes to the (leading-order) BK evolution, whereas its hard ($x\sim\order{1}$) part contributes to NLO. This ``unsubtracted'' formulation yields a positive-definite cross-section by construction, including when using different RC prescriptions in the impact factor and in the high-energy evolution, as above explained. This has been explicitly verified by the numerical calculations in \cite{Ducloue:2017mpb}, which also found that the NLO corrections are negative and significantly large --- they reduce the cross-section by up to 50\% w.r.t. the respective prediction at leading-order (LO).  A similar behavior has been recently observed in the NLO calculation of the DIS  structure functions \cite{Ducloue:2017ftk}: once again, the  ``unsubtracted''  version of the factorization leads to positive results.

Ref.~\cite{Ducloue:2017mpb} has also considered an alternative formulation of the NLO calculation, which differs from the ``canonical'' one in the presence of RC corrections, in order to study the scheme dependence of the results. In this new formulation, the one-loop calculation of the impact factor is fully rewritten in coordinate space and the transverse momentum of the produced parton is introduced via a final Fourier transform. In this context, it is possible and also natural to use a {\it coordinate-space} argument for the RC also in the impact factor, and thus have a unified treatment for the RC throughout the calculation. Clearly, the final results for the cross-section need not be exactly the same in this scheme and the ``canonical'' one (which uses a {\it momentum-space} RC prescription in the impact factor), yet they were expected to lie close to each other, because the Fourier transform roughly identifies momenta with the inverse of coordinates.  So, it came as a real surprise (and also as a bad news for the reliability of the NLO formalism as a whole) when it turned out that the numerical results are dramatically different within these two schemes: the NLO corrections obtained in the new scheme have the opposite sign as compared to those in the ``canonical'' scheme, and they are tremendously larger --- by one to two orders of magnitude, depending upon the final transverse momentum \cite{Ducloue:2017mpb}.

It is our main purpose in this paper to understand the origin of this puzzle and thus restore the predictive power of the NLO formalism. In our analysis, we shall trace this problem to the fact that the choice of a scale for the running of the coupling does not commute with the Fourier transform and show that this mathematical property is enough to explain the dramatic consequences observed in practice. This may look surprising in view of the fact that the scale dependence of the running coupling is quite weak --- merely logarithmic. Yet, as we shall show, it is precisely this logarithmic decrease of the RC with increasing transverse momenta (or decreasing transverse sizes) --- which is of course the expression of asymptotic freedom --- which is responsible for the problem identified in our NLO calculation. Namely, when used in coordinate space, the size-dependent RC may effectively act as a fictitious ``potential'', which allows for hard scattering (due its logarithmic singularity in the limit of a zero size) and hence can transfer a very high momentum to the produced parton even in the absence of any physical scattering. This fictitious scattering produces a spurious component in the particle production, which at high transverse momenta decreases slower than the physical tail and hence dominates over the latter.

The physical interpretation of this technical argument becomes more transparent if one starts in coordinate space. In order to produce a very hard parton, with a transverse momentum much larger than the saturation momentum of the target,  the produced particle must be accompanied by an unmeasured, recoil gluon, which carries the same transverse momentum but with the opposite sign. (It is precisely this recoil, or ``primary'' gluon, which is responsible for the one-loop contribution to the impact factor.) Soft primary emissions cannot contribute to the cross-section, simply by momentum conservation. But when the one-loop calculation is performed in coordinate space, this seemingly trivial kinematical constraint is built in a rather non-trivial way. The loop correction is then dominated by gluons which are emitted far away from their parent parton and which physically correspond to soft emissions. Yet, such soft gluons do not contribute to the production of that parent parton --- as expected from momentum conservation ---, because they are suppressed by the final Fourier transform. But this suppression can be spoilt by the scale-dependence of the RC: that is, even soft gluons can formally contribute to the production of a hard parton because the necessary transverse momentum can be (incorrectly) provided by the RC. This unphysical contribution is the source of the surprising results obtained with a coordinate-space prescription in Ref.~\cite{Ducloue:2017mpb}.

This discussion, developed in great detail in Sect.~\ref{sec:rcor}, shows that one must be very careful whenever the use of a RC is accompanied by a Fourier transform. There are other situations where the interplay between the running of the coupling and the Fourier transform has been reported to lead to difficulties --- e.g., to positivity violations for the ``unintegrated gluon distribution'' \cite{Giraud:2016lgg}, defined as a Fourier transform of the dipole scattering amplitude. This last problem will however not occur in our subsequent calculations. Presently, it is not clear to us whether these two problems are somehow related at a deeper level.

This being said, we shall nevertheless be able to identify at least one RC prescription which is formulated in coordinate space and which circumvents the aforementioned problem of the ``fake potential'' : this is the case where the argument of the RC is the transverse separation between the daughter gluons and their parent parton. This particular prescription quasi-commutes with the Fourier transform and in any case it cannot act as an unphysical source of transverse momentum. Indeed, we shall see that NLO calculations using this prescription gives results which are extremely close to those of the momentum-space RC prescription, thus confirming that the overall scheme-dependence of our NLO factorization is reasonably small. 

The daughter-gluon prescription\footnote{In the remaining part of this paper, we shall generally use the dipole picture of the BK evolution, as valid in the limit of a large number of colors; correspondingly,  we shall refer to this prescription as the ``daughter-dipole prescription''.} looks particularly appealing in that it can be simultaneously used within the one-loop calculation of the impact factor and the BK equation, thus providing a coherent scheme for the ensemble of the NLO calculation.\footnote{Note however a limitation of this scheme in practice, to be discussed in more detail in Sect.~\ref{sec:cf}: it cannot be used within the subclass of NLO corrections  (``the $C_{\rm F}$ terms'') that are concerned with the subtraction of collinear divergences. Indeed, this last operation has so far been performed only in momentum space \cite{Chirilli:2012jd}.} Results obtained with such a unified prescription will be presented as well and found to be rather close to those predicted by the ``canonical'' mixed-space prescription (momentum-space RC for the impact factor, but coordinate-space RC for the BK equation). However, one should keep in mind that the daughter-gluon prescription is not very well motivated in the context of the high-energy evolution --- especially for the asymmetric problems where there is a large disparity between the characteristic transverse sizes of the target and of the projectile (see the discussion in  Appendix~\ref{app:bal} below). This is in particular the case for the forward particle production in proton-nucleus collisions, especially when the produced parton is relatively hard. In such a case, we would still recommend the use of a ``mixed-space'' running-coupling prescription.

The paper is organised as follows. Sect.~\ref{sec:setup}  presents the NLO result for the quark multiplicity, as it will be used in this paper.  Besides introducing the notations and the relevant formulae, this presentation will also give us an opportunity to anticipate some of the subtleties with the use of a running coupling. Sect.~\ref{sec:rcor} is the main section of this paper. This is where we explain the origin of the puzzle with the coordinate-space RC prescription, as identified in Ref.~\cite{Ducloue:2017mpb}. We furthermore present an alternative RC prescription, still in coordinate space, which avoids this problem and can be meaningfully compared with the momentum-space prescription.  The discussion is illustrated with numerical results. Sect.~\ref{sec:cf} considers a special set of NLO corrections  (``the $C_{\rm F}$ terms'') which require a different treatment, since for them the transverse coordinate of the primary gluon has already been integrated out. (More technical aspects of this discussion are deferred to  Appendix~\ref{app:cf}.) Finally, Sect.~\ref{sec:conc} summarises our results and conclusions. For more clarity, we present in Appendix~\ref{app:bal} the usual RC prescriptions used in relation with the BK equation together with a physical discussion.

\section{\label{sec:setup} Single inclusive quark production at next-to-leading order}

In this section, we shall summarise the next-to-leading order results for the quark multiplicity in proton-nucleus collisions at forward rapidities together with the underlying physical picture. Our main focus will be on the ``unsubtracted'' formulation of the hybrid factorization to NLO \cite{Iancu:2016vyg}, but we shall also describe the original formulation by CXY \cite{Chirilli:2011km,Chirilli:2012jd} and the associated problems. Finally, we shall explain some subtleties which occur when trying to consistently include the running coupling effects in the various sectors of the calculation \cite{Iancu:2016vyg,Ducloue:2017mpb}.

\subsection{The hybrid factorization to NLO}

In the CGC effective theory \cite{Gelis:2010nm}, projectile partons are energetic enough so that their scattering off a large nucleus can be computed in the eikonal approximation:  the only effect of the scattering is a color rotation described by a Wilson line in the appropriate representation of the color group SU$(N_c)$. E.g., for a right-moving quark, the Wilson line reads
\beq
\label{wline}
V(\bx) = \rmP \left[\rmi g \int \dif x^+ A^-_a(x^+,\bx) t^a  \right],       
\eeq
where $g$ is the QCD coupling, $x^+$ the light-cone time of the projectile quark and $\bx$ its transverse coordinate, $A^-_a$ the relevant component of the color field of the nucleus and $t^a$ the  generators of SU($N_c$) in the fundamental representation.

We shall be interested in forward particle production in proton-nucleus collisions, and in particular in the quark multiplicity, i.e.~in produced quarks with transverse momentum $\bk$ and rapidity $\eta$ in the center of mass frame (COM). (Obviously, the respective gluon multiplicity will also contribute to the production of forward particles, but we shall not discuss its computation in the current work.) We choose the proton to be a right-mover, hence the nucleus to be left-mover, and work in the light-cone gauge $A^+_a=0$ in which the partonic structure of the proton (say, as encoded in the respective light-cone wavefunction) is manifest.  To compute the cross-section for quark production, we use the so-called ``hybrid factorization'' generally assumed to apply to such a ``dilute--dense'' scattering --- its validity has been so far demonstrated to next-to-leading order (NLO) in perturbation theory. In this factorization, the only information that we need to know about the dilute projectile (the proton) are the standard (``collinear'') parton distributions --- here, the quark distribution $x_p q(x_p, Q^2)$. In applications to phenomenology, one should also consider the fragmentation of the produced quark into hadrons, but here we shall ignore this final-state evolution, since irrelevant for the problems we would like to address.

Under these assumptions, the production of a quark at forward rapidity (i.e. for $\eta$ positive and large) and to leading order (LO) in pQCD is the result of the (multiple) scattering between a `valence' (in the sense of large $x_p$) quark from the incoming proton and the dense gluon distribution of the nucleus. The quark, originally collinear with the proton and hence with no initial transverse momentum, acquires its final transverse momentum $\bk$ through this scattering. The nuclear gluon distribution is {\it dense} since the typical gluons probed by this process have a very small value of the ``minus'' longitudinal momentum fraction $X\ll 1$ (see below) and relatively small transverse momenta, $q_\perp\sim Q_s(X)$, hence their occupation numbers are large. Here, $Q_s(X)$ is the nuclear saturation momentum at the longitudinal scale $X$ relevant for the scattering.

To the same accuracy, the scattering between the quark and the color fields representing the gluons in the nuclear target can be computed in the eikonal approximation, cf. \eqn{wline}. Then the LO quark multiplicity is computed as\footnote{The resolution scale ($Q^2$) dependence of the quark distribution will be omitted throughout this paper; in practice (and for the relatively hard transverse momenta of interest here: $k_\perp\gtrsim Q_s(X_g)$), one should choose $Q^2\sim k_\perp^2$.}
\begin{align}
	\label{lonc}
	\frac{\dif N^{{\rm LO}}}{\dif^2\bk\, \dif \eta} 
	=\, \frac{x_p q(x_p)}{(2\pi)^2}\,
 \mathcal{S}(\bk,X_g)\,,
\end{align}
where $ \mathcal{S}(\bk,X)$ is the ``unintegrated gluon distribution'' in the target, operationally defined as  the Fourier transform of the $S$-matrix  describing the elastic scattering between a small color dipole and the nucleus:
 \beq
 \label{sft}
 \mathcal{S}(\bk,X) = \int \dif^2\br\, 
 \rme^{-\rmi \bk \cdot \br}
 S(\br, X),
 \eeq
where in coordinate space we have
 \beq
 \label{scoord}
 S(\bx,\by;X) = \frac{1}{N_c}
 \left\langle \rmtr \left[V(\bx) V^{\dagger}(\by) \right]\right\rangle_{X}\,.
 \eeq
 The ``color dipole'' is made with a quark at $\bx$ and an antiquark at $\by$, which physically represent the produced quark in the direct amplitude and in the complex conjugate amplitude, respectively. 
In \eqn{sft}, $\br =\bx-\by$ is the dipole size and we also assume, for simplicity,  that the nucleus is quasi homogeneous in the transverse plane, so the $S$-matrix is independent of the impact parameter $\bm{b}=(\bx +\by)/2$.  The average in \eqn{scoord} is taken over the color field $A_a^{-}$ and with the CGC weight function which describes the high-energy evolution of our observable. Finally, the longitudinal momentum fractions for the quark, $x_p$, and for the gluons, $X_g$, which enter the formula \eqref{lonc} for the LO multiplicity are determined by the kinematics of the collision and of the produced quark, as follows:
\beq
\label{Xg}
x_p=\frac{k_\perp}{\sqrt{s}}\,\rme^\eta\,,\qquad X_g=\frac{k_\perp}{\sqrt{s}}\,\rme^{-\eta}=\frac{k_{\perp}^2}{x_p s}\,,
\eeq
where $k_{\perp}=|\bk|$ and $\sqrt{s}$ is the COM energy.  As anticipated, the forward kinematics ($\eta$ positive and large) corresponds to the regime $X_g\ll x_p < 1$.

Still at LO, one needs to resum to all orders the radiative corrections enhanced by powers of $\abar Y_g$, with $\abar\equiv \alpha_s N_c/\pi$ and $Y\equiv \ln(1/X_g)$, as generated by the high-energy evolution, i.e. by the successive emissions of soft gluons, which are strongly ordered in longitudinal momenta and whose effects should be computed in the eikonal approximation. To LO at least, this evolution can be associated with either the nucleus, or the incoming quark (or dipole), and in what follows we shall mostly adopt the first point of view (although the viewpoint of dipole evolution will be useful too for some arguments\footnote{This point of view is particularly useful in relation with the NLO impact factor, to be shortly discussed. The respective NLO correction has been computed \cite{Chirilli:2011km,Chirilli:2012jd} as a quantum fluctuation (the emission of the  ``primary gluon'') in the wavefunction of the incoming quark. Hence, the small-$x$ part of that emission was {\it a priori} interpreted as the first step in the (leading-order) high-energy evolution of the dipole projectile. But to the accuracy of interest, this can be alternatively viewed as the last step in the high-energy evolution of the nuclear target. See the discussion in Sect.~2.3 of Ref.~ \cite{Iancu:2016vyg} for details.}). $S(\bx,\by;X)$ is then the elastic amplitude for a ``bare'' dipole $(\bx,\by)$ which scatters off the nuclear gluon distribution evolved in $X$ from some initial value $X_0\sim 1$ (with $\abar\ln(1/X_0)\ll 1$) down to the scale $X_g\ll 1$ of interest. To LO in pQCD and in the limit of a large number of colors, $N_c\to\infty$, this evolution is described by the Balitsky-Kovchegov (BK) equation \cite{Balitsky:1995ub,Kovchegov:1999yj}, which reads
\begin{align}\label{BK}
X \frac{\del }{\del X} \,S(\bx,\by; X)=
 \frac{\abar}{2\pi}\, \int \rmd^2\bz\,
\frac{(\bx-\by)^2}{(\bx-\bz)^2(\bz-\by)^2}\,\Big[
 S(\bx,\bz; X) S(\bz, \by; X)
 -S(\bx,\by; X)\Big]\,,
 \end{align}
with $\bz$ denoting the transverse coordinate of the soft gluon emitted in one step of the evolution. The r.h.s. features the  ``dipole'' version of the LO BFKL kernel. The initial condition $S(\bx,\by;X_0)$ for this equation should be computed with a suitable model for the nuclear gluon distribution at $X_0$, like the McLerran-Venugopalan (MV) model \cite{McLerran:1993ni,McLerran:1993ka}.

At NLO, one needs to also include the ``pure-$\alpha_s$'' corrections, i.e. the radiative corrections of $\order{\alpha_s}$ which are not enhanced by $Y_g$. These can be divided into two classes: \texttt{(i)} NLO corrections to the high-energy evolution, i.e. to the kernel (more generally, to the structure) of the BK equation, and  \texttt{(ii)} NLO corrections to the ``impact factor'', i.e., to the ``hard'' matrix element which describes the quark-nucleus scattering in the absence of any evolution, meaning already for $X_g\sim X_0$. The LO impact factor describes the scattering between a bare quark collinear with the proton and the nucleus.
At NLO, the wavefunction of the incoming quark may also contain a gluon with ``plus'' longitudinal momentum fraction $x$ (w.r.t. the parent quark). Hence, the NLO impact factor must describe the scattering between this quark-gluon system and the nuclear target. So long as $x\ll 1$, this gluon emission can be computed in the eikonal approximation and its effect can then be included in the LO evolution. So, strictly speaking the NLO correction to the impact factor should only include relatively hard gluon emissions with $x\sim \order{1}$, which must now be computed {\it exactly} (i.e. beyond the eikonal approximation). In practice though, it turns out that separating the LO evolution from the NLO  correction to the impact factor is quite subtle (especially in the presence of running coupling corrections), as we shall see. For this reason, we shall prevent ourselves from doing such a separation and  mostly consider an ``unsubtracted'' expression for the NLO quark multiplicity  \cite{Iancu:2016vyg} in which the two types of corrections are still mixed with each other. 

As mentioned in the Introduction, our main goal in this analysis is to clarify the use of the running coupling in the calculation of the NLO multiplicity, notably in relation with the interplay between the high-energy evolution and the impact factor. For that purpose, we can ignore all the NLO corrections to the BK equation, except for those expressing the (one-loop) running of the coupling. That is, throughout this paper we shall use the running-coupling version of  \eqn{BK}, to be referred to as ``rcBK'', with various prescriptions for the argument of $\abar$, to be later specified.

The NLO expression for the quark multiplicity can be decomposed into two pieces, one proportional to the gluon Casimir $N_c$ and the other one to the quark Casimir $C_{\rm F}$ (see e.g. \cite{Ducloue:2016shw,Iancu:2016vyg} for details). In the formulation of Ref.~\cite{Iancu:2016vyg}, the first piece, proportional to $N_c$, is the one which encodes both the LO evolution and the part of the NLO impact factor which is most relevant for the present discussion. So, over most of the subsequent developments  we shall focus on this $N_c$-piece alone. (The $C_{\rm F}$-piece will be separately discussed, in Sect.~\ref{sec:cf}.) After including the NLO corrections proportional to $N_c$,  the quark multiplicity reads \cite{Iancu:2016vyg}
\begin{align}
	\label{nlonc}
	\frac{\dif N^{{\rm LO} + N_c}}{\dif^2\bk\, \dif \eta} 
	=\, &
	\frac{x_p q(x_p)}{(2\pi)^2}
	\mathcal{S}(\bk,X_0)	+
	\frac{1}{4\pi}\int_0^{1-X_g/X_0} \dif \xi\,
	\frac{1+\xi^2}{1-\xi}
	\nn
	& \times \left[\Theta(\xi-x_p)\frac{x_p}{\xi}\,
	q\left(\frac{x_p}{\xi}\right)
	\mathcal{J}(\bk,\xi,X(\xi)) - 
	x_p q(x_p) \mathcal{J}_v(\bk,\xi,X(\xi))\right].
\end{align}
The first piece in the r.h.s. is the would-be ``tree-level'' result, proportional to the LO impact factor $\mathcal{S}(\bk,X_0)$. The second term, expressed as an integral over $\xi$, encodes the dynamics of the (generally) \emph{non-eikonal} splitting of the incoming quark into a quark and the ``primary'' gluon, and the subsequent \emph{eikonal} scattering of this quark-gluon system off the evolved nucleus. The two terms under the integral, $ \mathcal{J}$ and $ \mathcal{J}_v$, refer to {\it real} and respectively {\it virtual} gluon emissions, where we recall that a parton is considered as ``real'' if it appears as an on-shell particle in the final state. (Some illustrative Feynman diagrams will be shown in Sect.~\ref{sec:rc},  in coordinate space.) 

The integration variable $\xi$ represents the splitting fraction of the quark, that is, $\xi= 1-x$, with $x$ the longitudinal momentum fraction of the primary gluon.  Since the ``plus'' component of the scattering partons is not modified in the eikonal approximation, also the observed quark carries a fraction $\xi$ of the parent quark. The latter has  therefore a longitudinal fraction $x_p/\xi$ wrt the projectile  proton.  On the other hand, in the  ``virtual'' term, which is needed for probability conservation, the final quark has the same longitudinal momentum fraction, equal to $x_p$, as the incoming quark. The fact that, when $\xi\ne 1$, the $ \mathcal{J}$ and $ \mathcal{J}_v$ terms in \eqn{nlonc} are weighted by {\it different} values of the  quark distribution function has therefore a trivial kinematical origin, but it will have important consequences in practice, as we shall see. Specifically, these two terms are given by \cite{Chirilli:2011km,Chirilli:2012jd}
 \begin{align}
 \label{J}
 \mathcal{J}(\bk,\xi,X(\xi)) &=
	\abar \int \frac{\dif^2\bq}{(2\pi)^2}
	\mathcal{S}(\bq, X(\xi))\bigg[
	\frac{2(\bk-\xi \bq)\!\cdot\!(\bk-\bq)}{(\bk-\xi \bq)^2(\bk-\bq)^2} 
	 - \int \frac{\dif^2\bel}{(2\pi)^2}
	\frac{2(\bk-\xi \bq)\!\cdot\!(\bk-\bel)}{(\bk-\xi \bq)^2(\bk-\bel)^2}
	\mathcal{S}(\bel, X(\xi))\bigg],
	\\*[0.2cm]
	\label{Jv}
\mathcal{J}_v(\bk,\xi,X(\xi)) &=
	\abar \int \frac{\dif^2\bq}{(2\pi)^2}
	\mathcal{S}(\bk, X(\xi))\bigg[
	\frac{2(\xi\bk-\bq)\!\cdot\!(\bk-\bq)}{(\xi\bk- \bq)^2(\bk-\bq)^2} 
	- \int \frac{\dif^2\bel}{(2\pi)^2}
	\frac{2(\xi\bk-\bq)\!\cdot\!(\bel-\bq)}{(\xi\bk-\bq)^2(\bel-\bq)^2}
	\mathcal{S}(\bel, X(\xi))\bigg].
 \end{align}
In writing these expressions, we have used the large-$N_c$ limit in order to factorize the dipole $S$-matrices appearing in the r.h.s. On the other hand, the color algebra associated with the emission of the primary gluon and with the subsequent scattering of the quark-gluon system has been computed exactly, for a generic value $N_c$. Hence, the overall factor $N_c$ (as encoded in the rescaled coupling constant $\abar= \alpha_s N_c/\pi$) can be trusted even beyond the large-$N_c$ limit. A similar discussion applies to the $C_{\rm F}$-terms to be presented in Sect.~\ref{sec:cf}.
 
Notice  that $\mathcal{J}$ and $\mathcal{J}_v$ have two sources of dependence upon $\xi$: one explicit in the transverse momentum kernels in Eqs.~\eqref{J}--\eqref{Jv}, which reflects the non-eikonal nature of the splitting, and one implicit in the rapidity arguments $X(\xi)$ of the dipole $S$-matrices, which expresses the evolution of the nuclear gluon distribution from the initial scale $X_0$ down to the longitudinal scale $X(\xi)$ probed by the effective projectile made with the quark and the primary gluon. This value $X(\xi)$ depends upon $\xi$ (and in general is larger than $X_g$) because the emission of the primary gluon ``consumes'' a rapidity interval $\Delta y=\ln(1/x)$, thus reducing the rapidity interval available to the evolution of the gluon distribution in the target, which now only goes up to
$Y_g-\Delta y= \ln(x/X_g)$. These considerations suggest that
  \beq
  \label{Xx}
  X(\xi) = \frac{X_g}{x} = \frac{X_g}{1-\xi} = \frac{k_{\perp}^2}{x_p s (1-\xi)}\, , 
  \eeq
 where the last equality follows after also using \eqn{Xg}. This turns out to be approximately correct, but only when the transverse momentum of the produced quark is large enough,  $k_{\perp} \gtrsim Q_s(X)$, with $Q_s(X)$ the target saturation momentum (see  \cite{Iancu:2016vyg} for details). This is indeed the most interesting kinematics for our purposes and the only to be considered in what follows.  Notice that the constraints $X_g \le X(\xi) \le X_0$ --- as coming from energy-momentum conservation together with our choice for the initial condition for the BK evolution --- imply indeed $0\le \xi \le 1-X_g/X_0$, in agreement with the integration limits visible  in \eqn{nlonc}.
 
As anticipated, the quantities  $\mathcal{J}$ and $\mathcal{J}_v$  include both the LO evolution (in the sense of rcBK) and the NLO corrections to the impact factor, albeit in a rather subtle way. The part of the LO evolution associated with the emission of a soft {\it primary} gluon is generated by the region $\xi\simeq 1$ (i.e. $x\ll 1$) of the integrals over $\xi$ in Eqs.~\eqref{J} and \eqref{Jv}; in this region, one can set $\xi=1$ within the emission kernels, which is tantamount to working in the eikonal approximation.  As for the subsequent, soft, gluon emissions responsible for the high-energy evolution over the remaining rapidity interval $\ln[X_0/X(\xi)]$, they are of course encoded in the various $S$-matrices, which are obtained by solving rcBK with initial condition ${S}(\br,X_0)$ and then making a Fourier transform to momentum space. Concerning the NLO corrections to the impact factor, these have two origins: the non-eikonal ($\xi\ne 1$ in the emission kernels) part of the primary gluon emission and the running coupling corrections to the $\abar$ prefactor visible in Eqs.~\eqref{J}--\eqref{Jv}, which so far has been formally treated as a fixed coupling. The separation of the LO evolution from the NLO impact factor together with various prescriptions for the running couplings (in the primary vertex and the BK equation) will be discussed in the next sections.

\subsection{\label{sec:sub} Unsubtracted, subtracted and CXY expressions}

In this section, we shall decompose the NLO result in \eqn{nlonc} into a leading order piece plus NLO corrections to the impact factor. Then we shall perform additional approximations in order to recover the original expression for the NLO quark multiplicity, by Chirilli, Xiao, and Yuan (CXY) \cite{Chirilli:2011km,Chirilli:2012jd}, which features the ``plus''-prescription in the integral over $\xi$. In turn, this  ``plus''-prescription is a variant of the ``$k_T$-factorization'', here applied to the unintegrated gluon distribution in the nucleus. Our main message from this discussion is that the ``$k_T$-factorization'', which is local in rapidity, is not equivalent to our general formula \eqref{nlonc}, but rather involves additional approximations which can be troublesome in practice. The subsequent manipulations are {\it a priori} valid at {\it fixed} coupling. Their extension to a running coupling is quite subtle and can bring additional complications, as we shall see in the next sections.

We start by introducing more compact notations, to be used only in this section. Namely, we
rewrite  \eqn{nlonc} as
 \beq
 \label{nlounsub}
 \frac{\dif N^{{\rm LO} + N_c}}{\dif^2\bk\, \dif \eta}  = 
 \frac{x_p q(x_p)}{(2\pi)^2}\,
 \mathcal{S}(\bk,X_0)
 +\int_0^{1-X_g/X_0} 
	\frac{\dif \xi}{1-\xi} \, \mathcal{K}(\bk,\xi,X(\xi))
 \equiv \frac{\dif N^{\rm IC}}{\dif^2\bk\, \dif \eta}
 +\frac{\dif N^{N_c,{\rm Unsub}}}{\dif^2\bk\, \dif \eta},
 \eeq
where the definition of $\mathcal{K}$ is easily understood by comparing to \eqn{nlonc}.
The first term in the r.h.s. is the tree-level contribution or equivalently the initial condition, while the superscript in the second term stands for ``unsubtracted'' and will be shortly explained.

The LO contribution to the quark multiplicity should be recovered from  \eqn{nlonc} in the limit where the primary gluon emission is evaluated in the eikonal approximation. As already explained, this corresponds to replacing $\xi\to 1$ within the kernels in Eqs.~\eqref{J}--\eqref{Jv}, as well as in the quark distribution accompanying the ``real'' emission, but not in the rapidity arguments $X(\xi)$ of the various $S$-matrices. This yields
\begin{align}
	\label{nlolo}
	\frac{\dif N^{{\rm LO}}}{\dif^2\bk\, \dif \eta} 
	=  &
	\frac{x_p q(x_p)}{(2\pi)^2}
	\mathcal{S}(\bk,X_0)	+
	\frac{x_p q(x_p)}{2\pi}\int_0^{1-X_g/X_0} 
	\frac{\dif \xi}{1-\xi} \,\big[
	\mathcal{J}(\bk,\xi=1,X(\xi)) - 
	 \mathcal{J}_v(\bk,\xi=1,X(\xi))\big].
\end{align}
To see that this reproduces indeed the expected LO result from \eqn{lonc}, we can use the identity
  \begin{align}
 \label{bkmom}
 \mathcal{S}(\bk,X_g) = 
 \mathcal{S}(\bk,X_0)
 +2 \pi \int_{X_g}^{X_0}
 \frac{\dif X}{X}\, 
 \big[
 \mathcal{J}(\bk,\xi=1,X) - \mathcal{J}_{v}(\bk,\xi=1,X) 
 \big],
 \end{align}
which is (in compact notations) the Fourier transform of the integral version of the LO BK equation \eqref{BK}.
 
 To separate leading from next-to-leading order contributions in   \eqn{nlonc}, we subtract the  LO result in the form of  \eqn{nlolo} and then add it back in its original form in  \eqn{lonc}. One thus finds
 \beq
 \label{nlosub}
 \frac{\dif N^{{\rm LO} + N_c}}{\dif^2\bk\, \dif \eta}  = 
 \frac{x_p q(x_p)}{(2\pi)^2}\,
 \mathcal{S}(\bk,X_g)
 +\int_0^{1-X_g/X_0} 
	\frac{\dif \xi}{1-\xi} \, \big[\mathcal{K}(\bk,\xi,X(\xi))
 -\mathcal{K}(\bk,\xi=1,X(\xi))\big]
 \equiv \frac{\dif N^{\rm LO}}{\dif^2\bk\, \dif \eta}
 +\frac{\dif N^{N_c,{\rm Sub}}}{\dif^2\bk\, \dif \eta},
 \eeq
 where we have used the compact notation introduced in \eqn{nlounsub}. Notice that the integrand in the ``subtracted'' piece has no singularity as $\xi\to 1$ (i.e. $x\to 0$), meaning that the integral over $\xi$ develops no ``small-$X_g$'' logarithm. This is in agreement with the fact that the longitudinal logarithm generated by a soft primary emission has been included in the evolution of the LO dipole $\mathcal{S}(\bk,X_g)$. Due to this subtraction, the integral in \eqn{nlosub} is a pure-$\alpha_s$ effect --- a NLO correction to the impact factor associated with a relatively hard primary emission.
 
Clearly,  Eqs.~\eqref{nlounsub} and \eqref{nlosub} are equivalent to each other, since related by exact manipulations. Notice however that in going from  \eqn{nlounsub} to \eqn{nlosub} one has added and subtracted a large term (the LO contribution) and in that process one has used the momentum-space version of the BK equation, \eqn{bkmom}. In other terms, the ``subtracted'' version,  \eqn{nlosub}, involves a fine cancellation between two large contributions and this cancellation works only so long as the dipole $S$-matrix obeys the LO BK equation. Any approximation or numerical error in the solution to the latter may spoil this cancellation and thus wash out the equivalence with the original, ``unsubtracted'', expression  \eqref{nlounsub}. Indeed, the numerical calculations in \cite{Ducloue:2017mpb} have confirmed the equivalence of the two forms in a wide range of transverse momenta. But at the same time they have shown that, due to numerical errors, some small oscillations persist for high transverse momenta when using the subtracted form. This means that for practical purposes, one is guaranteed to get more stable results when using the unsubtracted form of the quark multiplicity.  

Neither \eqn{nlounsub} nor \eqn{nlosub} correspond to the standard $k_{\rm T}$-factorization, since the dipole $S$-matrices are evaluated at the ``floating'' scale $X(\xi)$ given in \eqn{Xx}. In order to arrive at the $k_{\rm T}$-factorized formula presented in \cite{Chirilli:2011km,Chirilli:2012jd}, which is local in $X$, certain approximations need to be made. First, one observes that due to the subtraction in \eqn{nlosub}, the integral is dominated by small values $\xi\ll 1$. Hence, to the NLO accuracy of interest, it is justified to \texttt{(i)}  replace the rapidity argument of the dipole $S$-matrix  by its value at $\xi=0$, i.e.~$\mathcal{S}(\bk,X(\xi)) \to \mathcal{S}(\bk,X_g)$, and \texttt{(ii)}  neglect $X_g/X_0\ll 1$ in the upper limit of the integral over $\xi$. These approximations yield
 \beq
 \label{nlocxy}
 \frac{\dif N^{{\rm LO} + N_c}}{\dif^2\bk\, \dif \eta}\bigg|_{\rm CXY}  = 
 \frac{x_p q(x_p)}{(2\pi)^2}\,
 \mathcal{S}(\bk,X_g)
 + \int_{0}^1
 \frac{\dif \xi}{1-\xi} \, \big[\mathcal{K}(\bk,\xi,X_g)
 -\mathcal{K}(\bk,\xi=1,X_g)\big]
 \equiv \frac{\dif N^{\rm LO}}{\dif^2\bk\, \dif \eta}
 +\frac{\dif N^{N_c,{\rm Sub}}}{\dif^2\bk\, \dif \eta}
 \bigg|_{\rm CXY},
 \eeq
where all the $S$-matrices (explicit or implicit) in the r.h.s. are now evaluated at the rapidity scale $X_g$, that is, for the same rapidity separation from the target as that of the original quark.

\eqn{nlocxy} is not any more equivalent to Eqs.~\eqref{nlounsub} and \eqref{nlosub} and, despite the seemingly reasonable approximations, it is rather pathological as it rapidly becomes negative when increasing the transverse momentum of the produced quark (see e.g. the numerical results in  \cite{Ducloue:2017mpb}). The reason is that the replacement $X(\xi)\to X_g$ in the argument of the dipole $S$-matrix leads to an over-subtraction: the negative contribution proportional to $\mathcal{K}(\bk,\xi=1,X_g)$ becomes too large in magnitude and overcompensates for the LO piece in $\mathcal{S}(\bk,X_g)$. Moreover, the extension of the upper limit from $1-X_g/X_0$ to 1 is not physically motivated, since it violates constraints imposed from the correct kinematics, and thus it contains spurious contributions. 

\subsection{\label{sec:rc} The running coupling prescription: why is this a problem} 

As explained in the Introduction, the experience with the BFKL and BK equations demonstrates that the running coupling (RC) corrections are large for the high-energy evolution, to the point that they should better be viewed as a part of the LO formalism. There are several reasons for that. The non-locality of the BFKL/dipole kernels, which is further enhanced by the BFKL diffusion, implies that widely separated transverse (coordinate or momentum) scales, with {\it a priori} different values for the running coupling, can contribute to the evolution at a same, given, scale. Furthermore, the effects of the running are amplified by the evolution. For instance, the {\it saturation exponent} $\lambda_s\equiv \dif\ln Q_s^2/\dif Y$, which is a main prediction of the BK equation, is roughly twice to three times smaller when computed with a running coupling than with a fixed coupling. Accordingly, the evolutions of $Q_s$, hence of the gluon distribution, in the two scenarios --- fixed coupling vs. running coupling ---  deviate {\it exponentially} from each other with increasing $Y$. For such reasons, it is crucial to use the RC version of the BK equation \cite{Kovchegov:2006wf,Kovchegov:2006vj,Balitsky:2006wa,Balitsky:2008zza} when computing the quark multiplicity, both at leading-order, cf. \eqn{lonc}, and at next-to-leading order, cf. \eqn{nlonc}. This however introduces complications that we shall discuss in this and the next coming sections.

For consistency with the evolution equation, the explicit factor $\abar$  appearing in Eqs.~\eqref{J}--\eqref{Jv} for $\mathcal{J}$ and $\mathcal{J}_v$, which controls the emission of the primary gluon, must be treated as a {\it running} coupling too. Since Eqs.~\eqref{J}--\eqref{Jv}  are explicitly written in transverse momentum space,  it seems natural that the scale for the running of that coupling must be a suitable combination of the transverse momenta involved in the emission vertex.  When the  momentum $k_{\perp}$ of the outgoing quark is sufficiently hard, $k_\perp \gtrsim Q_s(X_g)$, one can simply chose $\abar(k_{\perp})$. This choice is also consistent with other kinematical approximations underlying Eqs.~\eqref{J}--\eqref{Jv}, notably the relation \eqref{Xx} for $X(\xi)$ (see  \cite{Iancu:2016vyg} for details). Conversely, the BK equation is generally formulated and solved in transverse coordinate space, cf. \eqn{BK}, where  it is more natural to choose a scale for the RC which is a combination of the dipole sizes involved in the splitting vertex (see below for some examples). Albeit perhaps unaesthetic, such a mixed choice for the RC prescriptions --- a momentum-space prescription in the calculation of $\mathcal{J}$ and $\mathcal{J}_v$, but a coordinate-space prescription for the high-energy evolution --- is not necessarily a problem: as we shall see, it still allows for meaningful results. However this has the drawback to introduce some conceptual ambiguities, which ultimately affect the predictive power of the NLO calculation. Specifically,  the following issues have been identified by previous studies \cite{Iancu:2016vyg,Ducloue:2017mpb}.

\texttt{(a)} The fact that the NLO result \eqref{nlonc} for the quark multiplicity reduces, as it should, to the respective LO result \eqref{lonc} when taking the eikonal limit for the primary emission relies in a crucial way on the {\it momentum-space} version of the LO BK equation, i.e.  \eqn{bkmom}.  But the Fourier transform and the choice of a prescription (scale) for the running of the coupling do {\it not} commute with each other, because of the scale dependence of the RC. That is, if one inserts a {\it momentum}-space RC in \eqn{bkmom}, like $\abar(k_\perp)$, then the result is not the same as the Fourier transform $\mathcal{S}(\bk,X_g)$ of the solution to the {\it coordinate}-space rcBK, which appears in \eqn{lonc}. As a consequence, there is a mismatch between the natural calculation of the quark multiplicity at LO, as based on  \eqn{lonc}, and the should-be ``LO limit'' of the corresponding expression at NLO. Such a mismatch is indeed visible in the numerical calculations in Ref.~\cite{Ducloue:2017mpb} and it is quite sizeable (the two predictions for the LO multiplicity can differ by up to 30\%; see also Fig.~\ref{fig:NLOsub}.A below).

\texttt{(b)} By the same argument as above, the equivalence between the ``unsubtracted'' and ``subtracted'' versions of the NLO result for the multiplicity, cf. Eqs.~\eqref{nlounsub} and \eqref{nlosub}, is violated when choosing different RC prescriptions for the primary vertex and the BK equation, respectively. In view of the fine-tuning inherent to the construction of the ``subtracted'' formula  \eqref{nlosub}, this mismatch may have important consequences. As argued in \cite{Iancu:2016vyg}, it may contribute to the ``negativity problem'' which afflicts the original calculation by CXY. Once again, this is confirmed by the numerical study in Ref.~\cite{Ducloue:2017mpb}. The calculation using the ``unsubtracted'' formula together with a mixed RC prescription leads to results which are stable and physically meaningful: as in the fixed coupling case, the quark multiplicity receives a negative correction at NLO level, but it remains positive and smooth for all final momenta $k_\perp$. On the contrary, the  ``subtracted''  formula leads to a pathological result, which suddenly turns negative at some intermediate value of $k_\perp$. This will be further discussed in
Sect.~\ref{sec:rcor} (see in particular Fig.~\ref{fig:NLOsub}.B).


\begin{figure*}[t]
\begin{center}
\begin{minipage}[b]{0.40\textwidth}
\begin{center}
\includegraphics[width=0.95\textwidth,angle=0 ]{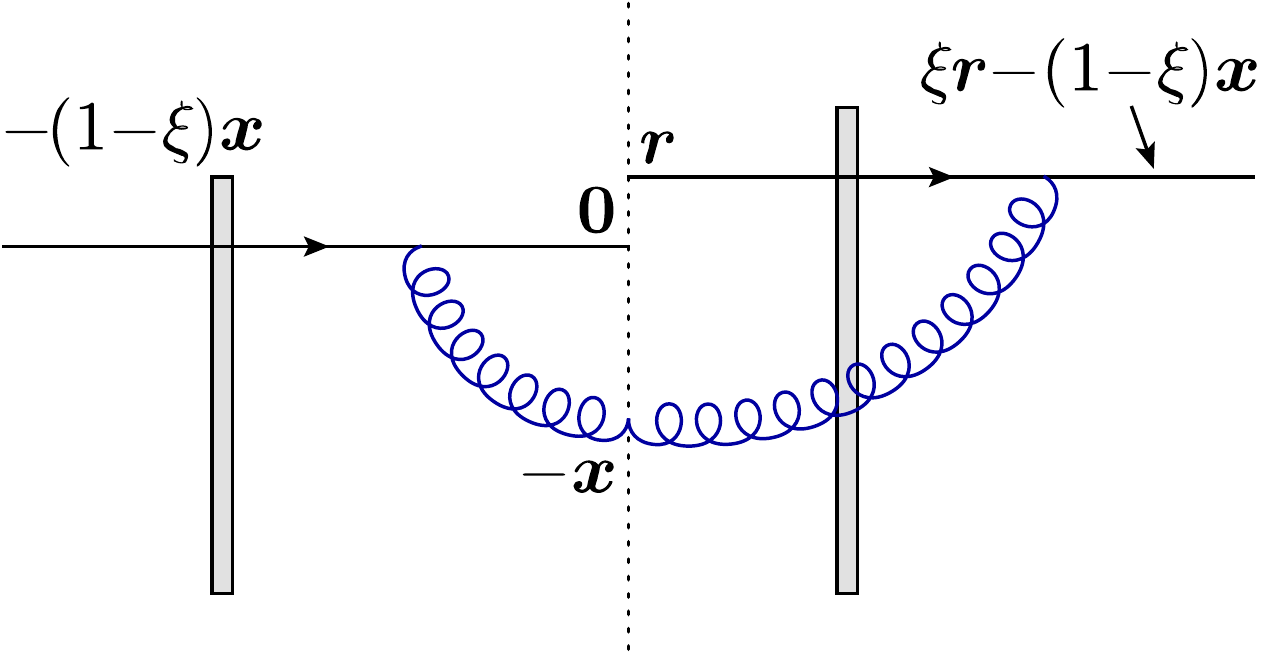}\\(A)\vspace{0cm}
\end{center}
\end{minipage}
\hspace{0.05\textwidth}
\begin{minipage}[b]{0.40\textwidth}
\begin{center}
\includegraphics[width=0.95\textwidth,angle=0]{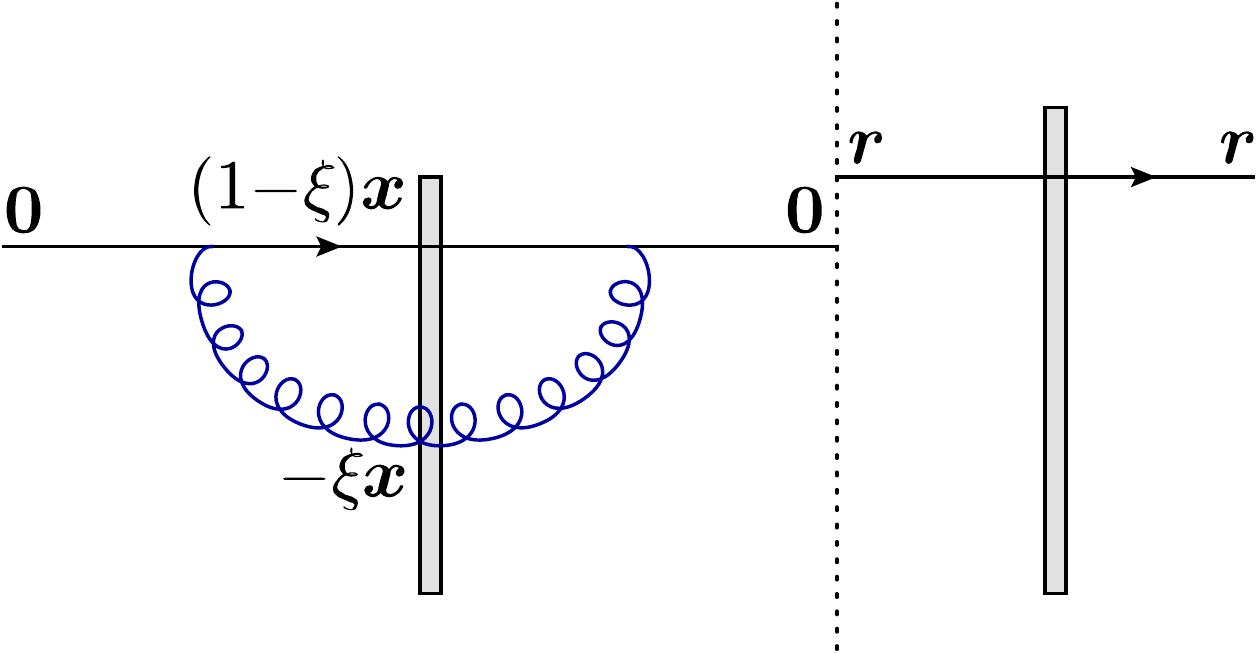}\\(B)\vspace{0cm}
\end{center}
\end{minipage}
\end{center}
\caption{\label{fig:Jco} \small Diagrams illustrating the real (A) and virtual (B) $N_c$-terms in the coordinate-space representation. The coordinate of a quark changes by the emission of the primary gluon (for $\xi\neq 1$), but not by the interactions with the target (represented by the thin vertical rectangles). Diagram (A) illustrates the second term in \eqn{Jco}, whereas diagram (B) corresponds to the second term in \eqn{Jcov}.}
\end{figure*}

These considerations motivated the recent proposal in Ref.~\cite{Ducloue:2017mpb} (see the Appendix there) to reformulate the calculation of the NLO multiplicity fully in coordinate space.  This should make it possible to use RC prescriptions which are consistent between the BK equation and the NLO impact factor and thus remove the ambiguities mentioned at points \texttt{(a)} and \texttt{(b)} above. Moreover, this would allow one to make contact with the phenomenology of DIS at HERA (e.g. in order to constrain the initial condition for the BK equation). Such a reformulation is indeed possible, at least for the $N_c$-terms under consideration (as we shall see, the situation is somewhat different for the $C_{\rm F}$-terms): indeed, Eqs.~\eqref{J}--\eqref{Jv} can be identically rewritten as the following Fourier transforms 
\begin{align}
    \hspace*{-0.7cm}
	\label{Jco}
	\mathcal{J}(\bk,\xi,X(\xi)) = &   
	\int\dif^2 \br\, \rme^{-\rmi \bk \cdot \br}
	J(\br,\xi,X(\xi))
	\nn
	& \hspace*{-1.6cm} \equiv  \int \dif^2 \br \,
	\rme^{-\rmi \bk \cdot \br}
	\int \frac{\dif^2 \bx}{(2\pi)^2}
	\,\abar\,
	\frac{2 \bx \!\cdot\! (\bx + \br)}{\bx^2(\bx + \br)^2}
	\left[ 
	S(\br + (1-\xi) \bx,X(\xi))
	-S(-\xi\bx,X(\xi)) S(\br + \bx,X(\xi))
	\right],
\end{align}
\begin{align}
\hspace*{-0.7cm}
	\label{Jcov}
	\mathcal{J}_v(\bk,\xi,X(\xi)) = &
	\int\dif^2 \br\, \rme^{-\rmi \bk \cdot \br}
	J_v(\br,\xi,X(\xi))
	\nn
	\equiv & \int \dif^2 \br \,
	\rme^{-\rmi \bk \cdot \br}
	\int \frac{\dif^2 \bx}{(2\pi)^2}
	\,\abar\,
	\frac{2}{\bx^2}
	\left[ 
	S(\br - (1-\xi) \bx,X(\xi))
	-S(-\bx,X(\xi)) S(\br + \xi \bx,X(\xi))
	\right],
\end{align}
where the coupling $\abar$ has been inserted {\it inside} the double integral over $\br$ and $\bx$, to emphasise that, when using a coordinate-space prescription, this coupling may depend upon all the  available transverse-coordinate scales and upon $\xi$. Some representative diagrams are shown in Fig.~\ref{fig:Jco}. Notice that the Fourier transforms leading from Eqs.~\eqref{J}--\eqref{Jv} to Eqs.~\eqref{Jco}--\eqref{Jcov} were {\it a priori} computed for a {\it fixed coupling}\,; the replacement of $\abar$ by a RC (with a coordinate-space prescription) must be done only {\it after} the Fourier transform.

One can similarly write the integral form of the coordinate-space BK equation \eqref{BK} as follows 
\begin{align}
 \label{bkco}
 S(\br,X_g)
 =\; &  S(\br,X_0)
 + \int\limits_{X_g}^{X_0}\frac{\dif X}{X}
 \int \frac{\dif^2 \bx}{2\pi}\,
 \abar(\br,\bx)
 \frac{\br^2}{\bx^2(\br+\bx)^2}
 \left[ S(-\bx,X) S(\br+\bx,X) - S(\br,X)\right]  \nn
  = \;&
 S(\br,X_0)
 +2 \pi \int\limits_{X_g}^{X_0}
 \frac{\dif X}{X}\, 
 \left[
 J(\br,\xi=1,X) - J_{v}(\br,\xi=1,X) 
 \right],
 \end{align}
where the rewriting in the second line emphasises that this is the same as  the Fourier transform of  \eqn{bkmom}. 

The explicit calculations of the one-loop RC corrections to the BK equation~\cite{Kovchegov:2006wf,Kovchegov:2006vj,Balitsky:2006wa,Balitsky:2008zza} show that these corrections are potentially large whenever there is a strong disparity between the sizes of the three dipoles (the parent and the two daughter ones) involved in the splitting. One can minimise these corrections by choosing the argument of the running coupling as the size of the {\it smallest} dipole: $\abar(\br,\bx) =\abar(r_{\rm min})$, with $r_{\rm min} \equiv {\rm min}\{|\br|, |\bx|,|\br-\bx|\}$. But when the dipole sizes are comparable with each other, there is some arbitrariness which is formally an NNLO effect. Besides the aforementioned {\it smallest dipole prescription}, one can also use other prescriptions, like the \emph{Balitsky prescription}~\cite{Balitsky:2006wa} or the \emph{fastest apparent convergence}~\cite{Iancu:2015joa}. Still, one should be rather careful with some choices: for example, for certain values of the dipole sizes the Balitsky prescription takes unphysically small values and one may question if the contribution from the respective phase space is properly computed (see the discussion in~\cite{Iancu:2015joa}). 

The coordinate-space prescriptions used in Eqs.~\eqref{Jco}--\eqref{Jcov} for the NLO impact factor and, respectively, the BK equation \eqref{bkco} become consistent with each other if the respective running couplings coincide with each other in the limit $\xi \to 1$. When this condition is satisfied, one recovers e.g. the equivalence between the ``unsubtracted'' and ``subtracted'' versions of the quark multiplicity at NLO.  This was indeed the case for the RC prescriptions (the Balitsky prescription and a certain generalisation of it for $\xi\neq 1$, whose precise form is not important for our purposes) used in the coordinate-space calculation presented in the Appendix of \cite{Ducloue:2017mpb}. However, the results of that particular calculation turned out to be extremely surprising (and in particular very different from those obtained within the same paper by using the momentum-space RC $\abar(k_\perp)$ within the ``unsubtracted'' scheme): the NLO corrections not only change sign (i.e.~they become positive), but they are also unacceptably large (by one to two orders of magnitude larger than the LO result, depending upon the value of $k_\perp$). Since this problem looks identical for both the  ``unsubtracted'' and the ``subtracted'' formulations of the NLO result, it is clear that its origin must be different from that of the ``negativity'' problem previously discussed. In the next sections, we shall clarify the origin of this problem and also propose a new RC prescription in coordinate space which avoids that problem and leads to physically meaningful results. Yet, as we shall also explain, that particular prescription is still not entirely satisfactory and cannot be viewed as the ultimate solution to the problem of writing the cross section fully in coordinate space.

\section{\label{sec:rcor} Coordinate-space prescriptions for the running coupling}
 
In this section, we shall explore in more detail the coordinate-space calculation of the NLO quark multiplicity based on Eqs.~\eqref{Jco}-\eqref{bkco}, with the purpose of clarifying the puzzle raised by the results in the Appendix of \cite{Ducloue:2017mpb}.  The problem becomes more severe --- in the sense that the deviation from the expected physical results (as obtained with momentum-space running coupling) becomes larger and larger --- when increasing $k_{\perp}$ (see Fig.~6 in Ref.~\cite{Ducloue:2017mpb} and also Fig.~\ref{fig:j-ratios-p} below). For that reason, in what follows we shall concentrate on the high-momentum tail of the quark distribution, at $k_{\perp} \gg Q_s$ (but our arguments will marginally hold down to $k_{\perp} \sim Q_s$). For such hard momenta, the exact form of the running coupling used in \cite{Ducloue:2017mpb} is not needed, and it just suffices to realise that it reduces to $\abar(r_{\perp})$. At a first glance, $r_{\perp}$ looks as a very reasonable length scale for the argument of the coupling: indeed, for $k_{\perp} \gg Q_s$ the momentum space choice is unambiguously $\abar(k_{\perp})$, and $\br$ is the variable conjugate to $\bk$. However, as we show in the following, the running coupling choice and the Fourier transform do not ``commute'', and therefore the choice $\abar(r_{\perp})$ turns out to be inappropriate.

\subsection{A fake potential}  

To illustrate the issue, let us start by exploring a simpler example,
which is still very close to our actual physical problem\footnote{In
  fact, it is very similar to one of the $C_{\rm F}$ terms discussed
  in section \ref{sec:cf}.}. Specifically, we consider the following two
  quantities
 \begin{align}
 	\label{nk}
 	&\mathcal{N}_k \equiv \abar(k_{\perp})\, 
 	\mathcal{S}(\bk)
 	=\abar(k_{\perp}) 
 	\int \dif^2 \br\, \rme^{-\rmi \bk \cdot \br} S(\br), 
 	\\
 	\label{nr}
 	&\mathcal{N}_r \equiv 
 	\int \dif^2 \br\, 
 	\abar(r_{\perp})
 	\rme^{-\rmi \bk \cdot \br} S(\br).
 \end{align}
 Note that both quantities are functions only of $\bk$, and the index
 just refers to the form of the coupling being used.
 We will will evaluate the above expressions using the MV model as an
 input, i.e.~with \beq
 \label{srmv}
 S(r_{\perp}) = \exp 
 \left( 
 -\frac{r_{\perp}^2 Q_s^2}{4}\,\ln \frac{1}{r_{\perp}^2 \Lambda^2}
 \right),
 \eeq 
where $Q_s$ is essentially\footnote{The saturation momentum is defined as the scale where the exponent in \eqn{srmv} becomes of order one. Hence, in the current notations, this scale should more properly read $Q_s^2 \ln (Q_s^2/\Lambda^2)$. This is indeed comparable with $Q_s^2$, which justifies our notations.} the target saturation momentum and $\Lambda$ is the usual QCD scale. It is implicitly assumed that the coordinate-space coupling in \eqn{nr} has been regularised in such a way to avoid the Landau pole at $r_{\perp} =1/ \Lambda$ and that the corresponding regularisation is smooth enough to avoid artifacts (like oscillations) when performing the Fourier transform. In practice, whenever an explicit regularisation is needed, we will use
\begin{equation}\label{eq:our-rc}
  \abar(r_{\perp}) = \frac{1}{2\bar b\left[\displaystyle 
 \ln\left(\frac{1}{r_{\perp} \Lambda}+ \rm{C}\right) \right]},
\end{equation}
where $\bar{b} = (11N_c - 2 N_{\rm f})/12 N_c$ and C a positive
constant (a free parameter to be varied in the numerical simulations)
that was introduced to ensure a smooth and regular behaviour of the RC
in the ``infrared'' (i.e. for large $r_\perp$). 

We consider the regime $k_{\perp} \gg Q_s$ and we first focus on the quantity $\mathcal{N}_k$ in \eqn{nk}. The result will be dominated by the single scattering term, since multiple scattering contributions will be suppressed by extra powers of $Q_s^2/k_{\perp}^2$. Notice that the zeroth order term in the expansion of the exponential in \eqn{srmv}, i.e. the unity (corresponding to no scattering), leads to a term proportional to $\delta(k_{\perp})$, which of course vanishes for our purposes. Keeping only the first order term in the expansion (i.e.~the contribution of a single scattering), we can write
 \beq
 \label{nkres}
 \mathcal{N}_k \simeq 
 \abar(k_{\perp}) \int \dif^2 \br\, \rme^{-\rmi \bk \cdot \br}
  \left( 
 -\frac{r_{\perp}^2 Q_s^2}{4}\,\ln \frac{1}{r_{\perp}^2 \Lambda^2}
 \right)
 =\frac{\abar(k_{\perp}) Q_s^2}{4}\,
 \nabla^2_{\bk} 
 \underbrace{\int \dif^2 \br\, \rme^{-\rmi \bk \cdot \br}
 \ln \frac{1}{r_{\perp}^2 \Lambda^2}}_{\displaystyle 4\pi/k_{\perp}^2} = \frac{4 \pi \abar(k_{\perp})Q_s^2}{k_{\perp}^4}.
 \eeq
This is the expected result for the high-$k_{\perp}$ tail of the distribution of a produced quark within the MV model, including the factor $\abar(k_{\perp})$ introduced by hand in the definition \eqref{nk}.

\begin{figure}[t] 
\centerline{\includegraphics[width=0.5\textwidth]{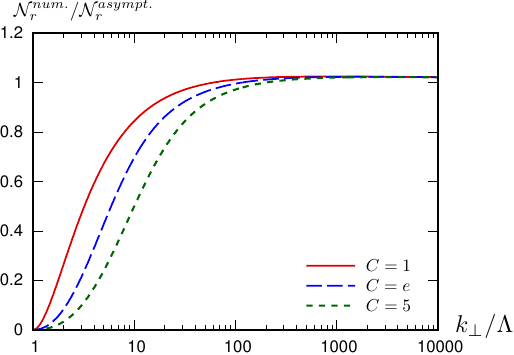}}
 \caption{\small The Fourier transform of the running coupling in \eqn{aft}, normalized to its asymptotic form as given analytically in \eqn{nrres}. We use the regularised version of the running coupling defined in \eqn{eq:our-rc}, with several choices for the constant C.}
 \label{fig:rc-reg}
\end{figure}

Now we move on to the hard-$k_{\perp}$ regime for the quantity
$\mathcal{N}_r$ in \eqn{nr} and, similarly to the above, one can check
that the single scattering contribution is again proportional to
$1/k_{\perp}^4$. However, this is not the dominant term anymore,
since, due to the inhomogeneity in $r_{\perp}$ of the coupling, the
unit piece of the $S$-matrix yields a non-zero result. Keeping only
that unit piece, our problem amounts to evaluating the Fourier
transform of the running coupling, that is,
%
%
\begin{align}
  \label{aft}
  \mathcal{N}_r 
  &\simeq \int \dif^2 \br\, \abar(r_{\perp})\rme^{-\rmi \bk \cdot \br} \\
  &\simeq \int \dif^2 \br\, \rme^{-\rmi \bk \cdot \br} 
    \frac{1}{\bar{b}\left(\displaystyle 
    \ln\frac{k_{\perp}^2}{\Lambda^2}+ 
    \ln\frac{1}{k_{\perp}^2 r_{\perp}^2}\right)}
    \simeq 
    \frac{1}{\bar{b}\ln (k_{\perp}^2/\Lambda^2)}
    \int \dif^2 \br\, \rme^{-\rmi \bk \cdot \br}
    \left[
    1 - \frac{\ln (1/r_{\perp}^2 k_{\perp}^2)}{\ln (k_{\perp}^2/\Lambda^2)}
    \right]
    \nonumber\\ 
  & = - \frac{1}{\bar{b}[\ln (k_{\perp}^2/\Lambda^2)]^2}
    \int \dif^2 \br\, \rme^{-\rmi \bk \cdot \br}
    \ln\frac{1}{r_{\perp}^2 k_{\perp}^2}.
  \label{nrexpand}
\end{align}
To write the second line above, we have used the fact that when
$k_{\perp}$ is very large, so is $1/r_{\perp} \sim k_{\perp}$, and we
can neglect the constant C next to the logarithm.
Strictly speaking,  the above integration should be restricted to
$\Lambda/k_{\perp} \ll k_{\perp} r_{\perp} \ll k_{\perp}/\Lambda$ for
the expansion to be valid. In practice though, the lower limit can be
safely set to 0 since $\Lambda/k_{\perp} \ll 1$ and the integrand
does not have a strong support when $r_{\perp}\to 0$, so the error
will be power-suppressed. Furthermore, the upper limit can be set to
infinity because now the integration is converging despite the fact
that we have ignored the regularisation constant C. We thus find
 \begin{equation}
 \label{nrres}
 \mathcal{N}_r \simeq - \frac{4 \pi}{\bar{b} [\ln (k_{\perp}^2/\Lambda^2)]^2}\,\frac{1}{k_{\perp}^2},
 \end{equation}
and we immediately notice that $\mathcal{N}_r$, when compared to the correct expression $\mathcal{N}_k$ given in \eqn{nkres}, is not only opposite in sign but also much larger in magnitude. (To verify the quality of the approximations leading from \eqref{aft} to \eqref{nrres}, we display in Fig.~\ref{fig:rc-reg} the ratio between the numerical result for the Fourier transform in \eqn{aft} and its analytic approximation in \eqn{nrres},  for several choices for the constant C in \eqn{eq:our-rc}.)
The result in \eqn{nrres} is very awkward, since it exhibits the power-law tail $\sim 1/k_{\perp}^2$ which emerges in the absence of an actual scattering and moreover dominates over the respective tail $\sim 1/k_{\perp}^4$ 
introduced by a single scattering (here treated in the MV model). In evaluating $\mathcal{N}_k$ the physical picture is indeed correct, as the transverse momentum $k_{\perp}$ has been provided (via a 2-gluon exchange) by the target potential. On the contrary, when evaluating $\mathcal{N}_r$ as shown in
Eqs.~\eqref{aft}--\eqref{nrres}, a hard momentum $k_{\perp}$ emerges due to the singular behavior of $\abar(r_{\perp})$ when $r_{\perp}\to 0$. In other words the running coupling is acting like a ``fictitious potential'' and this cannot be the correct physical picture.

\subsection{From parent dipole to daughter dipole prescriptions}

We now move to the actual problem under consideration, i.e. the
Fourier transforms which appear in the NLO calculation of the quark
multiplicity. As in the previous sections, we restrict to the
$N_c$-terms for the time being and postpone the discussion of the
$C_{\rm F}$-terms to the next section.
We therefore need to evaluate $\mathcal{J}(\bk,\xi)$ and
$\mathcal{J}_v(\bk,\xi)$ given by Eqs.~\eqref{Jco} and \eqref{Jcov}
respectively and we will focus on the real term.
For convenience, we will omit the rapidity arguments of $\mathcal{J}$,
$\mathcal{J}_v$ and $S$.

Before we discuss various RC prescriptions, let us establish what is
the range in transverse coordinate of the primary gluon, $\bx$,
which controls the integrations in \eqref{Jco}. Recall that we consider the production of a  relatively hard quark, with transverse momentum $k_{\perp}\gg Q_s$. Then the Fourier transform selects a relatively small value $r_\perp \sim 1/k_{\perp}\ll 1/Q_s$ for the parent dipole size. It is then easy to check that the dominant contribution to the integral over $\bx$ at a {\it fixed} (and small) value of $\br$ comes from relatively large daughter dipoles, with transverse sizes $x_\perp\simeq |\bx + \br| \gg r_\perp$. This is due to the behavior of the kernel in the respective integral, which vanishes as $x_\perp\to 0$ and decays only very slowly, as $1/x_\perp^2$, when $x_\perp\to\infty$. Of course, the integral will be eventually cutoff by the physics of saturation, i.e. by the fact that any of the $S$-matrices appearing in \eqn{Jco} will rapidly vanish when increasing $x_\perp$ above $1/Q_s$. But the dominant contribution comes from the range $r_\perp\ll x_\perp\lesssim 1/Q_s$ and more precisely from its upper limit, i.e. from $x_\perp\sim 1/Q_s$, due to the fact that the scattering amplitude $T(x_\perp)\equiv 1-S(x_\perp)$ grows quadratically with $x_{\perp}$ so long as $x_\perp\ll 1/Q_s$. We thus conclude that the would-be dominant contribution to the integral over $\bx$, as coming from very large daughter dipoles, is {\it independent} of $\br$ and hence it does not at all contribute to the subsequent Fourier transform:
\beq
 \label{jxggr}
 \mathcal{J}(\bk,\xi) \sim
 \frac{\abar}{2\pi^2}  
	\int \dif^2 \br \,
	\rme^{-\rmi \bk \cdot \br}
	\underbrace{\int_{r_{\perp}^2} \frac{\dif^2 \bx }{\bx^2}\,
 \left[ S((1-\xi) \bx) - S(-\xi\bx) S(\bx)
	\right]}_{\br-{\rm independent}} =0 	
	\quad \mbox{\rm for} \;\; x_{\perp} \gg r_{\perp}.
 \eeq
The latter is rather dominated by daughter dipoles whose sizes are comparable to that of their parent:
$x_\perp\sim r_\perp$.  For such small dipoles, one can use the single scattering approximation within the MV model, and thus deduce 
 \beq
 \label{jxsimr}
 \mathcal{J}(\bk,\xi) \sim 
	\frac{\abar(k_{\perp})}{\pi}  
	\int \dif^2 \br \,
	\rme^{-\rmi \bk \cdot \br}
	r_{\perp}^2 Q_s^2\, \ln \frac{1}{r_{\perp}^2 \Lambda^2}
	\sim \frac{\abar(k_{\perp}) Q_s^2}{k_{\perp}^4}
	\quad \mbox{\rm for} \;\; x_{\perp} \sim r_{\perp}, 
 \eeq
which is indeed a correction of  $\mathcal{O}(\abar)$ to the tree-level result at high-$k_{\perp}$, cf. \eqn{nkres}. A similar estimate holds for the virtual term $\mathcal{J}_v(\bk,\xi)$. Notice that some of the previous arguments have exploited the fact that $\xi\ne 1$, as is indeed the case for the generic values contributing to the integral over $x=1-\xi$ in \eqn{nlonc}. The limit $\xi\to 1$ of these results is well defined and is discussed below.

To summarise, the double integral in \eqn{Jco} is not controlled by the {\it typical} gluon emissions, which are favored by the emission kernel and are relatively soft ($p_\perp\sim Q_s$), but by the {\it rare} emission of a hard gluon with $p_{\perp}\sim k_{\perp}$.
This conclusion could have been anticipated on physical grounds: the relatively large momentum $k_{\perp}\gg Q_s$ of the measured quark cannot be provided by its scattering off the nuclear target, hence it must be balanced by the emission of an equally hard primary gluon, with transverse momentum  $p_{\perp}\simeq k_{\perp}$. In coordinate space, this corresponds indeed to a transverse size $x_\perp\sim r_\perp$. This correct physical picture should be preserved by any prescription used for the running of the coupling.

First of all, it is clear that the whole previous discussion remains unchanged if in Eqs.~\eqref{Jco} and \eqref{Jcov} one uses a running coupling with the scale set by the ``external''  momentum $ k_{\perp}$. This choice $\abar(k_{\perp})$ is physically well motivated when  $k_{\perp} \gg Q_s$, as already explained. So, in what follows we shall use the results computed with the momentum-space RC $\abar(k_{\perp})$ as a benchmark for analysing the suitability of other RC prescriptions formulated in coordinate space. For clarity, we denote the former with a superscript ``${mom}$'' and the latter with a superscript ``${pos}$''.

\setcounter{topnumber}{1}

\begin{figure*}[t]
\begin{center}
\begin{minipage}[b]{0.32\textwidth}
\begin{center}
\includegraphics[width=0.97\textwidth,angle=0]{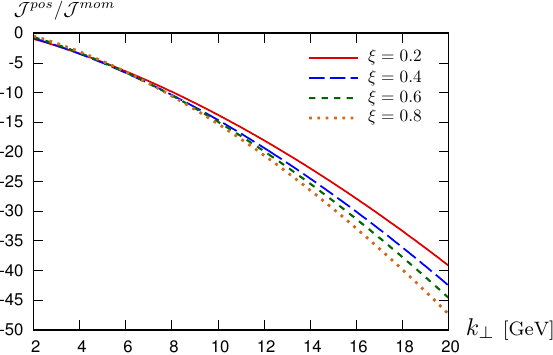}\\(A)\vspace{0cm}
\end{center}
\end{minipage}
\begin{minipage}[b]{0.32\textwidth}
\begin{center}
\includegraphics[width=0.97\textwidth,angle=0]{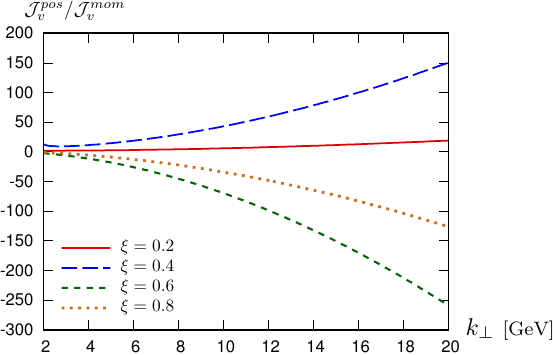}\\(B)\vspace{0cm}
\end{center}
\end{minipage}
\begin{minipage}[b]{0.32\textwidth}
\begin{center}
\includegraphics[width=0.97\textwidth,angle=0]{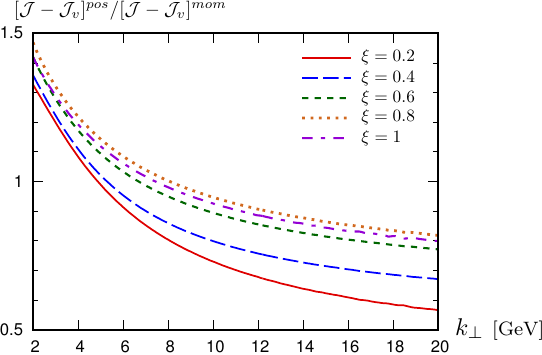}\\(C)\vspace{0cm}
\end{center}
\end{minipage}
\end{center}
\caption{\label{fig:j-ratios-p} \small The ``real'' and ``virtual'' $N_c$-terms,  $\mathcal{J}$ and $\mathcal{J}_v$, computed with the parent dipole prescription $\abar(r_{\perp})$ and normalised to the corresponding results obtained with the momentum-space prescription $\abar(k_{\perp})$.  Taken separately, the quantities $\mathcal{J}^{pos}$ and $\mathcal{J}_v^{pos}$ are pathological, in the sense of being  unphysically large and even having the wrong sign (see figs. A and B). However, their difference $\mathcal{J} -\mathcal{J}_v$ takes meaningful values (the ratio $|\mathcal{J} -\mathcal{J}_v|^{pos}/|\mathcal{J} -\mathcal{J}_v|^{mom}$ remains of $\order{1}$ when varying $k_\perp$ and $\xi$;  see fig. C), for the reasons explained around \eqn{deltaJ}.}
\end{figure*}

Since the Fourier transform in \eqn{Jco} selects parent dipole sizes
$r_\perp \sim 1/k_{\perp}$, it might seem reasonable to use the RC
$\abar(r_{\perp})$. This was essentially the choice made in
\cite{Ducloue:2017mpb} (at least for sufficiently high
$k_\perp$). Yet, as shown by the numerical results presented in the
Appendix of Ref.~\cite{Ducloue:2017mpb}, this prescription leads to
physically unacceptable results. The problem is further illustrated in
Figs.~\ref{fig:j-ratios-p}.A and B. Specifically,
Fig.~\ref{fig:j-ratios-p}.A shows the ratio
$\mathcal{J}^{pos}/\mathcal{J}^{mom}$ between the results for the
``real'' piece $\mathcal{J}$ obtained with the coordinate-space
prescription\footnote{For $\abar(r_{\perp})$, we used the
  regularisation shown in \eqn{eq:our-rc}, with the constant C
  chosen in such a way that $\abar(r_{\perp})\to 0.67$ as
  $r_{\perp}\to \infty$.} $\abar(r_{\perp})$ and the momentum-space
prescription
$\abar(k_{\perp})\equiv 1/[{\bar b}\ln(k_\perp^2/\Lambda^2)]$.  For
the dipole $S$-matrix, we have used the solution to rcBK with
smallest-dipole prescription $\abar(r_{\rm min})$ and with an initial
condition given by the MV model, evolved from $X_0=10^{-2}$ down to
$X=10^{-3}$. When evaluating $\mathcal{J}(\bk,\xi,X)$ according
\eqn{Jco}, we have treated $X$, $\xi$ and $k_\perp$ as independent
variables.\footnote{The actual dependence of $X$ on $\xi$ and
  $k_\perp$, shown in \eqn{Xx}, is unessential for the present
  argument.}  Similarly, Fig.~\ref{fig:j-ratios-p}.B shows the
corresponding ratio $\mathcal{J}_v^{pos}/\mathcal{J}_v^{mom}$ for the
``virtual'' piece $\mathcal{J}_v$. By inspection of these figures, it
is manifest that the results corresponding to the two RC prescriptions
are inconsistent. For instance, instead of being close
to one for sufficiently high $k_\perp$, the ratio
$\mathcal{J}^{pos}/\mathcal{J}^{mom}$ is negative and its modulus is
very large ($\gg 1$) and rapidly increasing with $k_\perp$, according
to a power law.

These results demonstrate the failure of the coordinate-space
prescription $\abar(r_{\perp})$ for the NLO impact factor. This
failure is easy to understand in light of our discussion in this and the previous 
sections. Recall that, by itself, the kernel in
\eqn{Jco} would favor large daughter dipoles with $x_\perp\gg r_\perp$
(meaning soft primary emissions), which however are eliminated by the
final Fourier transform $\br\to\bk$, since their contribution is
quasi-independent of $\br$, cf.~\eqref{jxggr}. However, the situation
changes in the presence of a RC like $\abar(r_{\perp})$ which
explicitly depends upon $r_\perp$: this coupling effectively acts as a
``potential'' which is singular as $r_{\perp}\to 0$, by asymptotic
freedom, meaning that it can transfer an arbitrarily hard transverse
momentum to the incoming quark. With the $\abar(r_\perp)$ prescription, 
the Fourier transform hence
introduces a spurious tail $\propto 1/k_\perp^2$ with a wrong sign,
cf.~\eqn{nrres}, which at large $k_\perp$ dominates over the physical
tail $\propto 1/k_\perp^4$ --- in agreement with the numerical results
exhibited in Figs.~\ref{fig:j-ratios-p}.A and B.

\begin{figure*}[t]
\begin{center}
\begin{minipage}[b]{0.32\textwidth}
\begin{center}
\includegraphics[width=0.97\textwidth,angle=0]{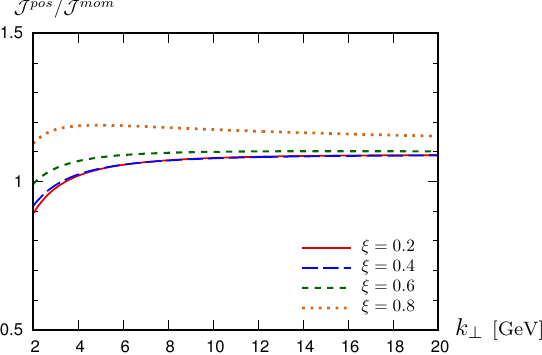}\\(A)\vspace{0cm}
\end{center}
\end{minipage}
\begin{minipage}[b]{0.32\textwidth}
\begin{center}
\includegraphics[width=0.97\textwidth,angle=0]{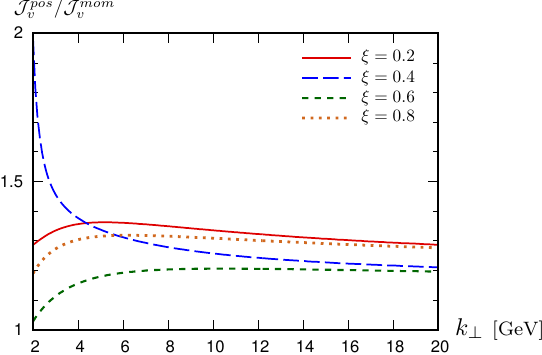}\\(B)\vspace{0cm}
\end{center}
\end{minipage}
\begin{minipage}[b]{0.32\textwidth}
\begin{center}
\includegraphics[width=0.97\textwidth,angle=0]{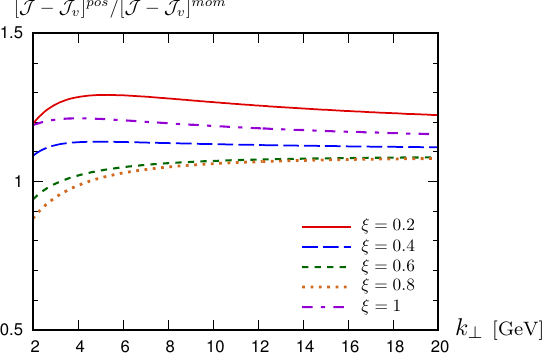}\\(C)\vspace{0cm}
\end{center}
\end{minipage}
\end{center}
\caption{\label{fig:j-ratios-d} \small The same ratios as in Fig.~\ref{fig:j-ratios-p} except that the coordinate-space prescription used for the RC coupling which controls the primary emission is the daughter dipole prescription $\abar(x_{\perp})$. All the ratios are now rather close to one (at least for sufficiently large values of $k_\perp$, where the present approximations are supposed to apply).}
\end{figure*}

At this stage, one may wonder about the implications of our previous
discussion for the solution to rcBK, where it is quite common to use a
RC prescription --- like $\abar(r_{\rm min})$ or the Balitsky
prescription discussed in Appendix~\ref{app:bal} --- which becomes
equivalent to the parent-dipole prescription $\abar(r_{\perp})$
whenever $r_\perp$ is the smallest scale. In that context, such a prescription is well-known to be well-behaved.
We show now that this is not in contradiction with our above arguments
that $\mathcal{J}$ and $\mathcal{J}_v$ are ill-behaved when computed with the
$\abar(r_{\perp})$ prescription.
 The crucial point is that the r.h.s. of the BK equation \eqref{bkco}
involves the {\it difference}  $\mathcal{J}(\bk,\xi, X)- \mathcal{J}_v(\bk,\xi, X)$ with $\xi =1$,
that is \beq
 \label{deltaJ}
 \mathcal{J}(\bk,\xi=1) - \mathcal{J}_v(\bk,\xi=1) 
 = \int \dif^2 \br\, 
 	\abar(r_{\perp})\,
 	\rme^{-\rmi \bk \cdot \br}
 	 \int \frac{\dif^2 \bx}{(2\pi)^2}
	\frac{\br^2}{\bx^2(\bx + \br)^2}
	\left[S(-\bx) S(\br + \bx) - S(\br) \right],
 \eeq  
 where we have directly inserted a RC $\abar(r_{\perp})$ and we
have omitted the rapidity arguments for simplicity.
The main difference compared to the previous discussion is that now
the unphysical contributions due to large dipoles cancel out in this
difference. Indeed, \eqn{deltaJ} involves the {\it dipole kernel}, which
decays much faster --- namely, as $r_\perp^2/x_\perp^4$ --- at large
$x_\perp\gg r_\perp$.  Using again the MV model for the dipole
$S$-matrix and restricting ourselves to the single scattering
approximation, one finds
 \begin{align}
 \label{deltaJss}
 \mathcal{J}(\bk,\xi=1) - \mathcal{J}_v(\bk,\xi=1) 
 &\simeq -
 \frac{Q_s^2}{8\pi}
 \int \dif^2 \br\,
 	\abar(r_{\perp})\,
 	\rme^{-\rmi \bk \cdot \br} r_{\perp}^2
 	 \int\limits_{r_{\perp}^2}^{1/Q_s^2} \frac{\dif x_{\perp}^2}{x_{\perp}^2}\,
 	 \ln \frac{1}{x_{\perp}^2\Lambda^2}
 	 \nn
 	 & \simeq \frac{Q_s^2}{16\pi}
 	 \int \dif^2 \br\,
 	\abar(r_{\perp})\,
 	\rme^{-\rmi \bk \cdot \br} r_{\perp}^2 
 	\ln^2 \frac{1}{r_{\perp}^2\Lambda^2}\,,
 \end{align}
where we have kept only the contribution from relatively large daughter dipoles, within the range $r_\perp\ll x_\perp \ll 1/Q_s$. (This is indeed the dominant contribution, since it is enhanced by an additional transverse logarithm.) The integrand in \eqn{deltaJss} is very different from that occurring in \eqn{jxggr} for the ``real'' piece $\mathcal{J}$  alone.  Compared to \eqn{jxggr},  the integration over $\bx$ now introduces an extra {\it logarithmic} dependence on the parent dipole size $r_\perp$, on top on that encoded in the RC $\abar(r_{\perp})$. This additional dependence is the actual physical source for the transverse momentum $\bk$ of the produced quark. The final Fourier transform in \eqn{deltaJss} is in fact similar to that appearing in \eqn{jxsimr} and thus yields a high-momentum tail $\sim 1/k_{\perp}^4$, as expected for the first step in the BK evolution of the quark multiplicity.

Albeit the previous argument was directly constructed for the limit $\xi\to 1$, as relevant for the BK equation, it actually holds for {\it generic} values of $\xi$ : the difference $\mathcal{J}(\bk,\xi, X)- \mathcal{J}_v(\bk,\xi, X)$ is free of the ``fake potential'' problem for any value $\xi\le 1$. This can be easily verified by making a  change of variables in \eqn{Jcov} in order to match the arguments of the various $S$-matrices there with those in \eqn{deltaJ}.  This is further confirmed by the numerical results in Fig.~\ref{fig:j-ratios-p}.C.  
This observation also shows that the ultimate reason why the calculation of the NLO multiplicity using the parent dipole prescription  $\abar(r_{\perp})$ is so ill behaved is because the ``real'' and ``virtual'' contributions to \eqn{lonc} are differently weighted by the quark distribution, which is taken at a longitudinal momentum fraction equal to $x_p/\xi$ in the ``real'' contribution, but equal to $x_p$ in the ``virtual'' one.

The previous discussion immediately suggests a better RC prescription to
be used in coordinate space: the scale of the RC should rather be set
by the {\it daughter} dipole size $x_\perp$. Indeed, with the choice
$\abar(x_{\perp})$, the spurious contribution generated by large
($x_\perp\gg r_\perp$) daughter dipoles remains independent of $\br$,
meaning it is eventually removed by the Fourier transform, as it should. Then the net
contribution comes solely from the physical configurations with
$x_\perp\sim r_\perp$. For such configurations, the coordinate-space
prescription $\abar(x_{\perp})$ should be equivalent to the
momentum-space one $\abar(k_{\perp})$. This is confirmed by the
numerical results shown in Figs.~\ref{fig:j-ratios-d}.A and B, which
are the counterpart of those in Figs.~\ref{fig:j-ratios-p}.A and B,
but with a daughter-dipole prescription $\abar(x_{\perp})$ in
coordinate space. As visible in these figures, the results are now
nicely consistent between the coordinate-space and the momentum-space
prescriptions. The remaining difference should be viewed as a measure
of the RC scheme-dependence of our calculation, and this dependence
turns out to be quite small (and decreasing with increasing
$k_{\perp}$, as expected).
Similarly, the behaviour of $\mathcal{J}-\mathcal{J}_v$, shown in
Fig.~\ref{fig:j-ratios-d}.C for the daughter-dipole prescription
 $\abar(x_{\perp})$, is also well-behaved and within our
 scheme-dependence.

\subsection{NLO results with various running coupling prescriptions}

\begin{figure*}[t]
\begin{center}
\begin{minipage}[b]{0.47\textwidth}
\begin{center}
\includegraphics[width=0.95\textwidth,angle=0]{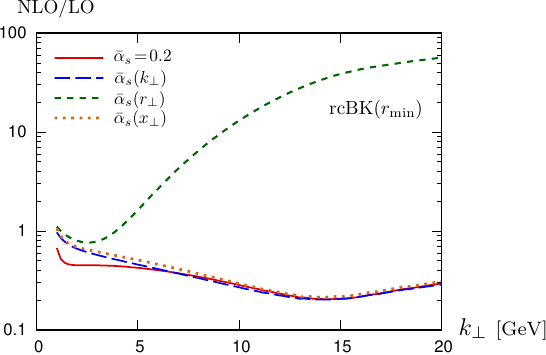}\\(A)\vspace{0cm}
\end{center}
\end{minipage}
\begin{minipage}[b]{0.47\textwidth}
\begin{center}
\includegraphics[width=0.95\textwidth,angle=0]{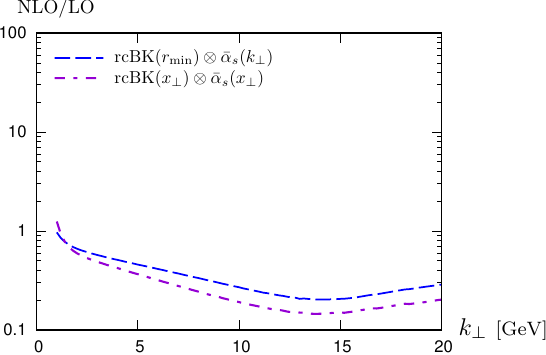}\\(B)\vspace{0cm}
\end{center}
\end{minipage}
\end{center}
\caption{\label{fig:NLOratio} \small Left: The ratio between the NLO cross-section \eqref{nlounsub} and the LO one, \eqn{lonc}, for different prescriptions for the running of the coupling which controls the emission of the primary gluon, as indicated in the legend of the figure. The evolution of the dipole $S$-matrix is always computed from the rcBK equation with smallest-dipole prescription  $\abar(r_{\rm min})$. More details on the kinematics are presented in the text. Right: The curve ``$\abar(k_{\perp})$'' is the same as in the left figure. The curve ``$\abar(x_{\perp})$'' is obtained by using the daughter dipole prescription both in the primary vertex and in the BK equation.}
\end{figure*}

\begin{figure*}[t]
\begin{center}
\begin{minipage}[b]{0.47\textwidth}
\begin{center}
\includegraphics[width=0.95\textwidth,angle=0]{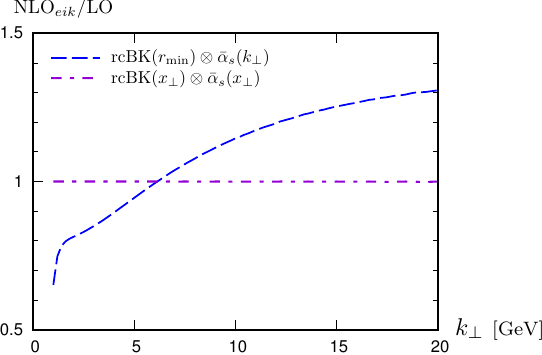}\\(A)\vspace{0cm}
\end{center}
\end{minipage}
\begin{minipage}[b]{0.47\textwidth}
\begin{center}
\includegraphics[width=0.95\textwidth,angle=0]{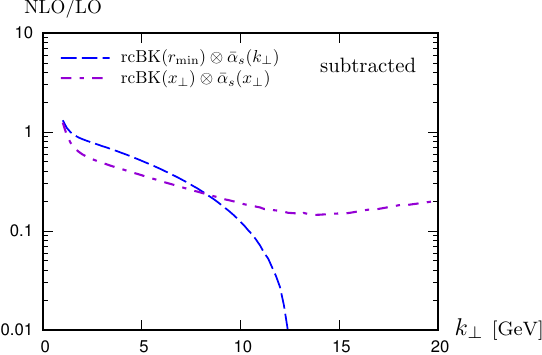}\\(B)\vspace{0cm}
\end{center}
\end{minipage}
\end{center}
\caption{\label{fig:NLOsub} \small Left: The ratio NLO$_{eik}$/LO between the eikonal limit of the NLO cross-section,  obtained by using $\mathcal{K}(\bk,\xi=1,X(\xi))$  in the integrand of \eqn{nlounsub}, and the LO result  \eqref{lonc}. The prescriptions used for the RC are exactly the same as in Fig.~\ref{fig:NLOratio}.B. Right: the predictions of the ``subtracted'' scheme, \eqn{nlosub}, for the same RC prescriptions as in the left figure: the curve ``$\abar(k_{\perp})$'' is pathologic (it shows a sudden drop at intermediate values of $k_\perp$), whereas the curve ``$\abar(x_{\perp})$''  coincides with the respective ``unsubtracted'' result in Fig.~\ref{fig:NLOratio}.B.
}
\end{figure*}

To further illustrate our discussion in this section, we exhibit in Figs.~\ref{fig:NLOratio} and \ref{fig:NLOsub} numerical results for the NLO multiplicity obtained with different prescriptions for the RC.  For more clarity, the NLO results are normalised by the respective LO predictions, as obtained by using \eqn{lonc} together with the solution to the BK equation with {\it running coupling}. The RC prescription for rcBK is the minimal dipole prescription $\abar(r_{\rm min})$, unless otherwise specified. As before, we use the MV model as an initial condition at $X_0=10^{-2}$ and present the results for the cross-section as a function of $k_\perp$ corresponding to a COM energy $\sqrt{s}=500~$GeV and a pseudo-rapidity $\eta=3.2$ for the produced quark. The corresponding values for $x_p$ and $X_g$ can be read off \eqn{Xg}.

All the curves presented  in Fig.~\ref{fig:NLOratio} are obtained by using the ``unsubtracted'' formula \eqn{nlounsub}. The 4 curves in Fig.~\ref{fig:NLOratio}.A  correspond to 4 different choices for the coupling which is explicit in the NLO impact factor  (i.e. the coupling which controls the emission of the primary gluon and whose running has been the main issue of this section):
 \texttt{(i)} a fixed coupling $\abar=0.2$ ,  \texttt{(ii)} a one-loop running coupling with momentum-space prescription,  $\abar(k_{\perp})$, \texttt{(iii)} a one-loop RC with parent dipole prescription,  $\abar(r_{\perp})$,  and finally \texttt{(iv)} a one-loop RC with daughter dipole prescription,  $\abar(x_{\perp})$. The regularisation used for the one-loop RC is as shown in \eqn{eq:our-rc}, with the constant C chosen in such a way that $\abar(r_{\perp})\to 0.67$ as $r_{\perp}\to \infty$. We observe that the 2 curves corresponding to $\abar(k_{\perp})$ and to $\abar(x_{\perp})$ are not only very close to each other (thus confirming the little scheme dependence in our calculation), but also very close to the corresponding results at fixed coupling. That is, the effect of the running of the coupling is not so important in so far as the primary emission is concerned. This is due to the fact that the running of the coupling is merely logarithmic and the interval in $k_{\perp}$ that we consider in Fig.~\ref{fig:NLOratio} is quite small. Besides, the effects of this particular coupling are not amplified by the evolution, unlike what happens for the coupling occurring in the BK equation.  On the other hand, the results obtained with the parent-dipole prescription $\abar(r_{\perp})$ are dramatically different, in agreement with the previous discussion in this section (and also with the original results in \cite{Ducloue:2017mpb}): they differ from the correct results by up to two orders of magnitude, and this difference keeps increasing with $k_\perp$, due to the spurious high-$k_\perp$ tail introduced by the Fourier transform of $\abar(r_{\perp})$.
 
The curve labelled as ``$\abar(k_{\perp})$'' in Fig.~\ref{fig:NLOratio}.B is exactly the same as the respective curve in Fig.~\ref{fig:NLOratio}.A. But the other curve in Fig.~\ref{fig:NLOratio}.B, denoted as  ``$\abar(x_{\perp})$'', is now obtained by systematically using the daughter dipole prescription $\abar(x_{\perp})$ throughout the calculation, {\it including} within rcBK. That is, it differs from the respective curve in Fig.~\ref{fig:NLOratio}.A by the RC prescription used for the dipole evolution. The two curves in Fig.~\ref{fig:NLOratio}.B are quite close to each other, albeit their difference is somewhat larger than that observed between the 3 ``physical'' curves in Fig.~\ref{fig:NLOratio}.A [notice however the different vertical scales used in these 2 figures]. Such a difference was in fact to be expected: a RC with daughter dipole prescription gives a stronger weighting to the emission of gluons with soft transverse momenta (i.e. to large daughter dipoles) and thus leads to a somewhat faster evolution. We therefore expect the difference between the 2 curves  in Fig.~\ref{fig:NLOratio}.B to grow with increasing the phase-space for energy evolution.

Having a unique RC prescription in the whole NLO calculation looks conceptually appealing, in that it renders the calculation more homogeneous.  In particular, this removes both ambiguities mentioned at points \texttt{(a)} and \texttt{(b)} in Sect.~\ref{sec:rc}, as numerically demonstrated in Fig.~\ref{fig:NLOsub}. The curves ``$\abar(k_{\perp})$'' and ``$\abar(x_{\perp})$'' shown in this figure are obtained by using the same RC prescriptions as in Fig.~\ref{fig:NLOratio}.B. As expected, the curve ``$\abar(x_{\perp})$'' which uses a unique, daughter-dipole, prescription in all the stages of the calculation, has an unambiguous LO limit (cf. Fig.~\ref{fig:NLOsub}.A) and yields identical results with both the ``unsubtracted'' and the ``subtracted'' formulations of the NLO multiplicity (cf. Fig.~\ref{fig:NLOsub}.B) --- at variance with the curve labelled  ``$\abar(k_{\perp})$'', which uses a mixed set of RC prescriptions.

This being said, the calculation underlying the curve ``$\abar(k_{\perp})$'' in Fig.~\ref{fig:NLOratio}.B, which uses a  mixed RC prescription together with the ``unsubtracted'' formula \eqref{nlounsub} for the cross-section, is still the one to be {\it a priori} trusted at high energy. This is so since, strictly speaking, the daughter dipole prescription $\abar(x_{\perp})$ is not fully suitable for the BK equation, in that it artificially accelerates the evolution. To understand that, recall that the dominant dipole configurations are not the same for the calculation of the NLO impact factor (i.e. for the emission of the primary gluon) and for the BK evolution (i.e. for the subsequent emissions of soft gluons, as resummed by the BK equation). Indeed, unlike the primary gluon, which is as hard as the produced quark (so the associated ``daughter dipole'' has a small size $x_\perp\sim r_\perp$, as previously discussed), the subsequent emissions in the ``hard-to-soft'' BK evolution involve predominantly gluons with smaller and smaller transverse momenta, or  ``daughter dipoles'' with larger and larger transverse sizes. For such typical emissions, pQCD (and in particular the experience with the DGLAP equation \cite{Dokshitzer:1991wu}) instructs us that the proper scale to be used in the RC is the transverse size of the {\it parent} dipole (see also the discussion in Appendix~\ref{app:bal} below). By asymptotic freedom, this typically leads to a smaller value for the coupling as compared to the daughter dipole prescription, and hence to a somewhat slower evolution. A further argument in favor of the momentum-space prescription $\abar(k_{\perp})$ will be provided by the discussion of the $C_{\rm F}$ terms in the next section.

\section{\label{sec:cf} The $C_{\rm F}$ terms}

In this section, we shall extend the previous discussion to the remaining contributions to the quark multiplicity at NLO, those proportional to the quark Casimir $C_{\rm F}$.  If we preferred to discuss these terms separately, it is not because they were exempted of the ``fake potential'' problem aforementioned --- as we shall see, they suffer from a similar problem when computed in coordinate space and with parent dipole running-coupling prescription ---, but rather because they encode a different physical regime. Unlike the $N_c$ terms, which include the small-$x$ limit of the primary gluon emission and thus overlap with the LO BK evolution, these $C_{\rm F}$ terms include the respective {\it collinear} limits --- i.e. the limits where the primary gluon is collinear with either the incoming quark, or with the outgoing one --- and hence overlap with the DGLAP evolution of the quark distribution function, and of the quark-into-hadron fragmentation function, respectively. These collinear limits signal themselves via (infrared) logarithmic divergences, which need to be regularised and subtracted away (since the respective contributions have already been included via the above mentioned DGLAP evolution). In practice, this is conveniently done by using dimensional regularisation together with minimal subtraction. The remaining, finite, terms are pure NLO corrections to the impact factor. This renormalisation procedure is explained in detail in Ref.~\cite{Chirilli:2011km,Chirilli:2012jd}, from which we shall simply quote the necessary results.

\begin{figure*}[t]
\begin{center}
\begin{minipage}[b]{0.32\textwidth}
\begin{center}
\includegraphics[width=0.95\textwidth,angle=0 ]{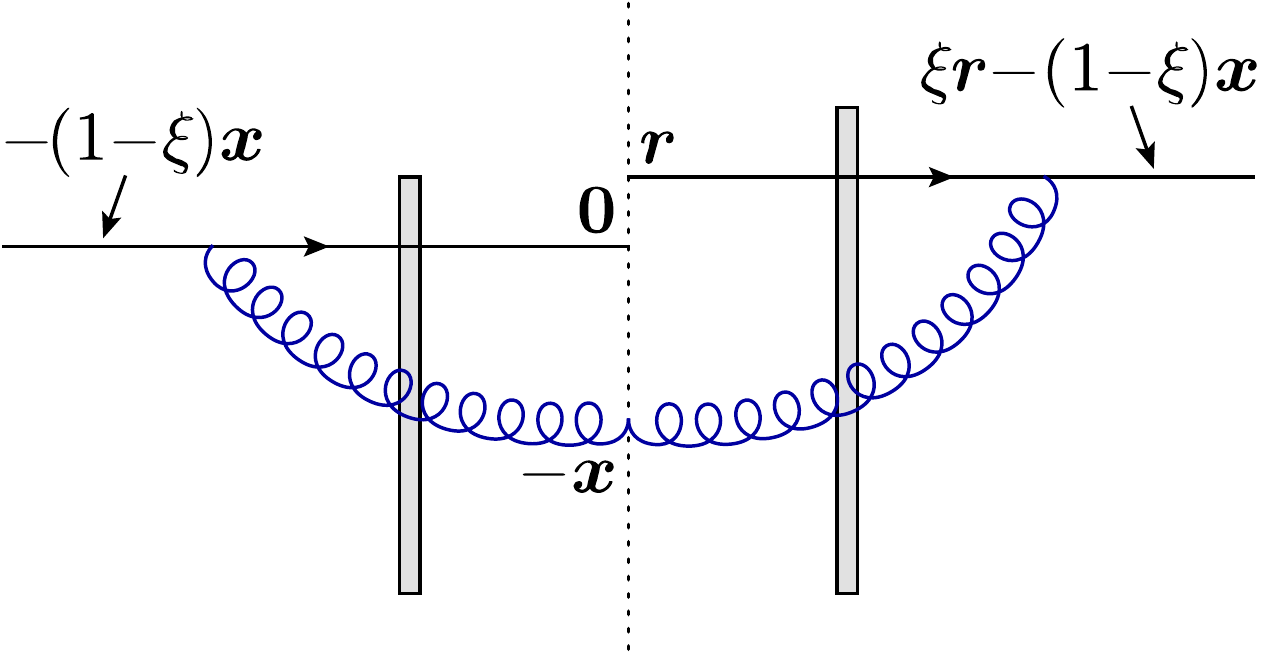}\\(A)\vspace{0cm}
\end{center}
\end{minipage}
\begin{minipage}[b]{0.32\textwidth}
\begin{center}
\includegraphics[width=0.95\textwidth,angle=0]{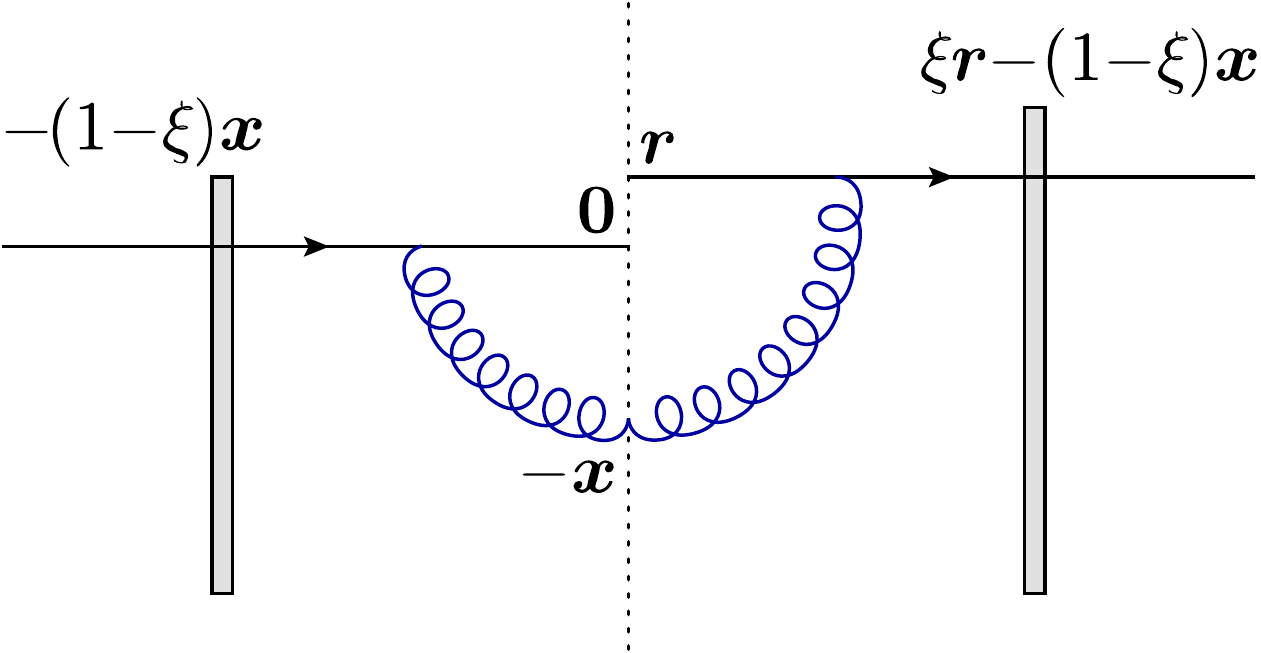}\\(B)\vspace{0cm}
\end{center}
\end{minipage}
\begin{minipage}[b]{0.32\textwidth}
\begin{center}
\includegraphics[width=0.95\textwidth,angle=0]{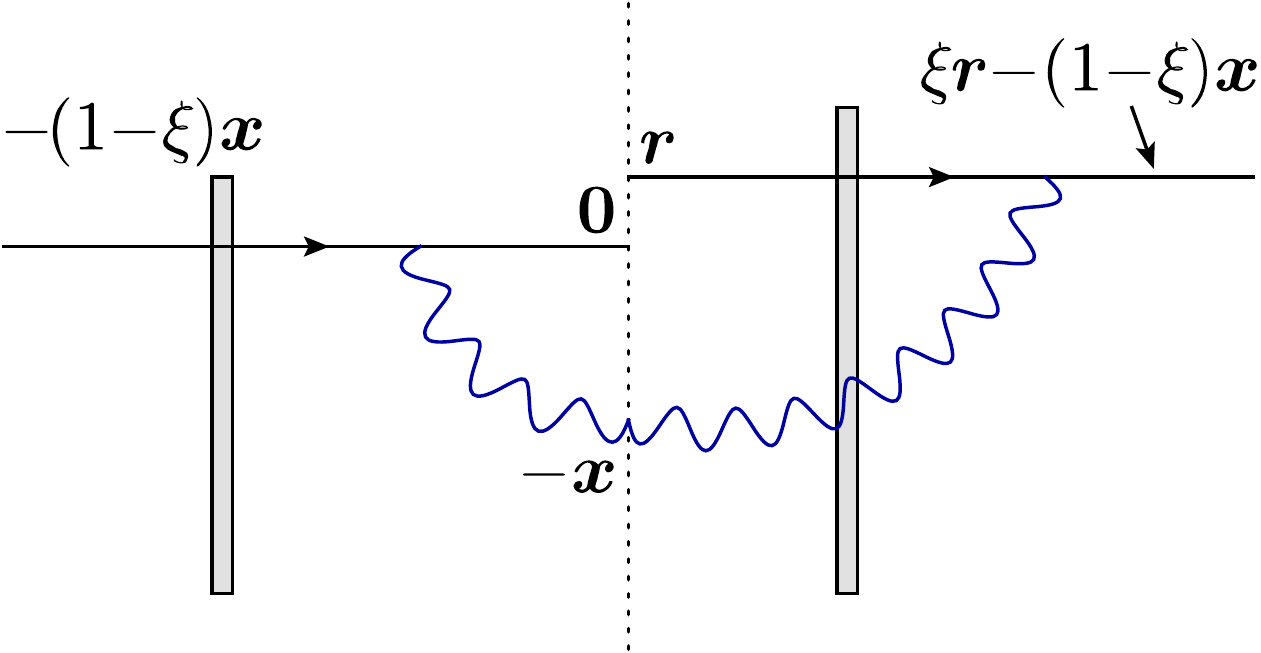}\\(C)\vspace{0cm}
\end{center}
\end{minipage}
\end{center}
\caption{\label{fig:Ico} \small Feynman graphs  contributing to the $C_{\rm F}$-term $\mathcal{I}$ in coordinate space, which illustrate the three terms in the brackets in \eqn{Ico}. In graph (A), the interactions between the gluon and the shockwave fully cancel between the direct amplitude and the complex conjugate amplitude, by unitarity: $V(-\bx) V^{\dagger}(-\bx)=1$ (here, the Wilson line is in the adjoint representation). The ``photon-like'' line replacing the primary gluon in graph (C) represents the {\it color singlet} piece of that gluon (hence it does not interact with the shockwave). Indeed, graph (C) represents only a {\it piece} of an actual Feynman graph, namely that piece of the diagram in Fig.~\ref{fig:Jco}.A which is suppressed at large $N_c$ (the singlet piece in the color decomposition $[N_c^2-1]=[N_c\times N_c] - [1]$ of the gluon).  Following Ref.~\cite{Chirilli:2011km,Chirilli:2012jd}, the overall factor  $1/2N_c$ multiplying  this {\it suppressed} piece has been conveniently rewritten as $1/2N_c= N_c/2 - C_{\rm F} $, i.e. as the difference between two {\it unsuppressed} contributions. This rewritting generates 2 terms, one proportional to $N_c/2$ that was included in the $N_c$-terms (as the first term in the r.h.s. of \eqn{J} for $ \mathcal{J}$), the other one proportional to $C_{\rm F}$, which is the third term in \eqn{Ico}.}
\end{figure*}

The contribution of the  $C_{\rm F}$ terms to the quark multiplicity at NLO can be written as
\begin{align}
	\label{nlocf}
	\frac{\dif N^{C_{\rm F}}}{\dif^2\bk\, \dif \eta} 
	= \frac{C_{\rm F}}{2 \pi N_c}
	\!\int_0^{1-X_g/X_0} \dif \xi\,
	\frac{1+\xi^2}{1-\xi}
	 \left[\Theta(\xi-x_p)\frac{x_p}{\xi}\,
	q\left(\frac{x_p}{\xi}\right)
	\mathcal{I}(\bk,\xi,X(\xi)) - 
	x_p q(x_p) \mathcal{I}_v(\bk,\xi,X(\xi))\right],
\end{align}
where $\mathcal{I}$ and $\mathcal{I}_v$ are ``real'' and ``virtual'' terms associated with the same process as described before: the non-eikonal splitting $q \to q g$ followed by the scattering off the nucleus. For more clarity, we first present the original version of these terms, prior to renormalization. When computed in transverse momentum space, they read 
\begin{align}
\label{I}
	\mathcal{I}(\bk,\xi,X(\xi)) &=
	\abar \int \frac{\dif^2\bq}{(2\pi)^2}
	\left[\frac{\bk-\bq}{(\bk-\bq)^2}
	- \frac{\bk-\xi \bq}{(\bk-\xi \bq)^2} \right]^2 
	\mathcal{S}(\bq, X(\xi)),
	\\
	\label{Iv}
    \mathcal{I}_v(\bk,\xi,X(\xi)) &=
	\abar \int \frac{\dif^2\bq}{(2\pi)^2}
	\left[\frac{\bk-\bq}{(\bk-\bq)^2}
	- \frac{\xi \bk-\bq}{(\xi\bk-\bq)^2} \right]^2 
	\mathcal{S}(\bk, X(\xi)).
\end{align}
The collinear divergences (at $\bq=\bk$ and $\bq=\bk/\xi$ in the ``real'' term and, respectively, 
at $\bq=\bk$ and $\bq=\xi\bk$ in the ``virtual'' term) are indeed manifest. It is a straightforward exercise in Fourier transforms to write the above formulae as double integrations in coordinate space, so that they look more similar in structure to those in Eqs.~\eqref{Jco} and \eqref{Jcov}. For example, for the ``real'' term we find
\begin{align}
\hspace*{-0.4cm}
\label{Ico}
	\mathcal{I}(\bk,\xi,X(\xi)) =
	\abar \int \dif^2\br\, \rme^{-\rmi \bk \cdot \br} 
	\int \frac{\dif^2\bx}{(2\pi)^2}
	\frac{\bx \cdot (\bx+\br)}{\bx^2 (\bx+\br)^2}
	\left[ S\big(\br, X(\xi)\big) 
	+S\big(\xi\br, X(\xi)\big) 
	-2S\big(\xi\br - (1-\xi)\bx, X(\xi)\big)\right],
\end{align}
where the three terms within the square brackets correspond to the terms obtained by expanding the square in the integrand of \eqn{I} and also to the three diagrams shown in Fig.~\ref{fig:Ico}.

The first two diagrams,  Figs.~\ref{fig:Ico}.A and B, are the expected counterparts of the two diagrams shown in Fig.~\ref{fig:Jco}, corresponding to the other possible time-orderings for the emission of the primary gluon. In the eikonal limit $\xi\to 1$ these four diagrams would together generate the first step in the BK evolution. This last statement also shows that, by themselves, the first two terms within the brackets in \eqn{Ico} --- corresponding to the squared terms, $1/(\bk-\bq)^2$ and $1/(\bk-\xi \bq)^2$, in \eqn{I} --- would also generate a {\it longitudinal} logarithm via the $\xi\to 1$ limit of the integral in \eqn{nlocf}; that is, they would contribute to the high-energy evolution as well. The third term in \eqn{Ico} --- corresponding to the ``crossed'' term in the square in \eqn{I} --- has been introduced by hand in Ref.~\cite{Chirilli:2011km,Chirilli:2012jd} in order  to subtract the  $\xi\to 1$ limit of the first two terms. More precisely, this term has been {\it added} to the $N_c$ terms and {\it subtracted} from the $C_{\rm F}$ terms, in such a way to disentangle the soft from the collinear divergences. The apparent mismatch between the color factors of the ``added'' and ``subtracted'' terms, $N_c$ and respectively $C_{\rm F}/2$, is irrelevant in the large $N_c$ limit of interest here. But even for finite $N_c$, this mismatch is precisely compensated  by that piece of the diagram in Fig.~\ref{fig:Jco}.A which would be suppressed at large $N_c$. This piece is illustrated in Fig.~\ref{fig:Ico}.C (see the caption to Fig.~\ref{fig:Ico} for more details). A similar discussion applies to the ``virtual'' term $\mathcal{I}_v$.

To summarise, the ``real'' and ``virtual'' $C_{\rm F}$-terms,  $\mathcal{I}$ and $\mathcal{I}_v$, separately vanish in the eikonal limit  $\xi \to 1$, thus guaranteeing that the integration in \eqn{nlocf} will not generate any large longitudinal logarithm. They also vanish  --- once again, due to the subtraction of the third term in \eqn{Ico} --- in the absence of any scattering, i.e~when $S(\br)=1$ or, equivalently, $\mathcal{S}(\bk)=(2\pi)^2\delta^{(2)}(\bk)$. A similar property holds for the $N_c$-terms, as manifest e.g. in Eqs.~\eqref{Jco}--\eqref{Jcov}. This property is important since it guarantees that there is no particle production in the absence of scattering, as expected on physical grounds.

However, Eqs.~\eqref{I}, \eqref{Iv} and \eqref{Ico} are afflicted with infrared divergences associated with collinear emissions. In momentum-space, these divergences were already noticed after Eqs.~\eqref{I}--\eqref{Iv}. Importantly, they refer to the squared terms {\it alone} (in either  $\mathcal{I}$ and $\mathcal{I}_v$); the respective crossed terms are free of singularities for any $\xi< 1$. In coordinate space, e.g.~in \eqn{Ico}, the infinities occur in the limit where the gluon is emitted very far from the quark, that is when $x_{\perp}$ gets very large. Once again, this happens only for the first two terms in the square bracket in \eqn{Ico}, corresponding to the first two diagrams in Fig.~\ref{fig:Ico}. In Ref.~\cite{Chirilli:2011km,Chirilli:2012jd}, the subtraction of the collinear divergences has been performed by working in momentum space, i.e. at the level of Eqs.~\eqref{I}--\eqref{Iv}, and for a {\it fixed} coupling. The final results after subtraction can be written as
\beq
\label{Ifin}
	\mathcal{I}^{\rm fin}(\bk,\xi,X(\xi)) =
	\abar\! \int\! \frac{\dif^2\br}{4\pi}\,
	S(\br, X(\xi))\ln \frac{c_0^2}{\br^2 \mu^2}
	\left(\rme^{-\rmi \bk \cdot \br} 
	+ \frac{1}{\xi^2} \rme^{-\rmi {\textstyle\frac{\bk}{\xi}} \cdot \br} 
	\right)
	- 2 \abar \!\int\! \frac{\dif^2 \bq}{(2\pi)^2}
	\frac{(\bk - \xi \bq)\cdot(\bk-\bq)}{(\bk - \xi \bq)^2(\bk-\bq)^2}
	\mathcal{S}(\bq, X(\xi)),
\eeq
\beq
\hspace*{-0.75cm}	
	\label{Ivfin}
    \mathcal{I}_v^{\rm fin}(\bk,\xi,X(\xi)) =
	\abar
	\left[\ln\frac{\bk^2}{\mu^2} + \ln(1-\xi)^2 \right]
	\frac{\mathcal{S}(\bk, X(\xi))}{2\pi}
	=
	\abar \left[\ln\frac{\bk^2}{\mu^2} + \ln(1-\xi)^2 \right]
	\int \dif^2\br\,
	\frac{S(\br, X(\xi))}{2\pi}\,
	\rme^{-\rmi \bk \cdot \br}	,
\eeq
with $\mu$ a factorization scale and $c_0 = 2 \rme^{-\gamma_\ssE}$.

\begin{figure*}[t]
\begin{center}
\begin{minipage}[b]{0.32\textwidth}
\begin{center}
\includegraphics[width=0.97\textwidth,angle=0]{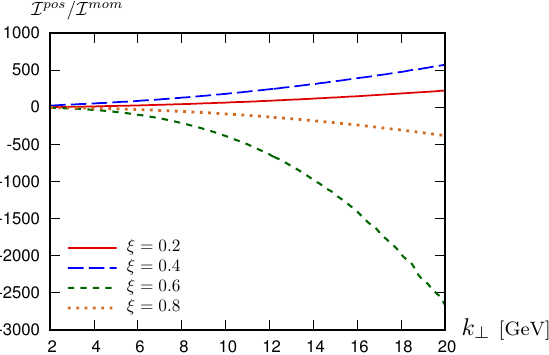}\\(A)\vspace{0cm}
\end{center}
\end{minipage}
\begin{minipage}[b]{0.32\textwidth}
\begin{center}
\includegraphics[width=0.97\textwidth,angle=0]{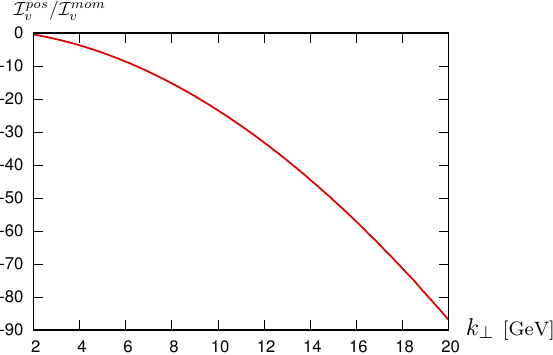}\\(B)\vspace{0cm}
\end{center}
\end{minipage}
\begin{minipage}[b]{0.32\textwidth}
\begin{center}
\includegraphics[width=0.97\textwidth,angle=0]{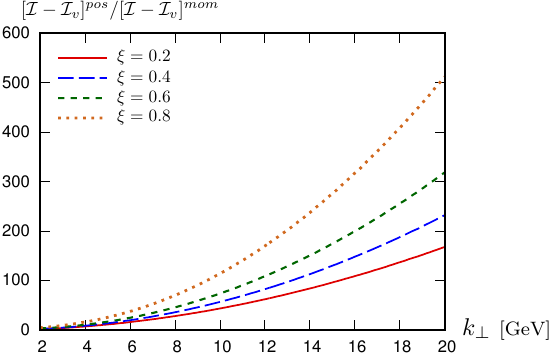}\\(C)\vspace{0cm}
\end{center}
\end{minipage}
\end{center}
\caption{\label{fig:i-ratios-d} \small The ``real'' and ``virtual'' $C_{\rm F}$-terms,  $\mathcal{I}$ and $\mathcal{I}_v$, computed with the parent dipole prescription $\abar(r_{\perp})$ and normalised to the corresponding results obtained with the momentum-space prescription $\abar(k_{\perp})$. In the middle graph, the values of $\xi$ need not be specified, since the ratio $\mathcal{I}_v^{pos}/\mathcal{I}_v^{mom}$ is independent of $\xi$, as manifest by inspection \eqn{Ivfin}. Also, we exclude the limiting value $\xi=1$ from these plots because the densities $\mathcal{I}$ and $\mathcal{I}_v$ are separately divergent in the limit $\xi\to 1$. Moreover, the difference  $\mathcal{I}^{mom}-\mathcal{I}_v^{mom}$ vanishes as $\xi\to 1$, as shown in Appendix \ref{app:cf}.}
\end{figure*}

Besides being finite, the above expressions for  $\mathcal{I}^{\rm fin}$ and $\mathcal{I}_v^{\rm fin}$ preserve the ``good'' properties of their un-regularised versions $\mathcal{I}$ and $\mathcal{I}_v$  that we previously discussed. First, they vanish in the absence of any scattering and for $\bk\neq \bm{0}$; this is obvious for $\mathcal{I}_v^{\rm fin}$ and can also be checked for $\mathcal{I}^{\rm fin}$. (When $S(\br)=1$ and $\mathcal{S}(\bq)=(2\pi)^2\delta^{(2)}(\bq)$, we can easily perform the integrations in \eqn{Ifin} and thus find that $\mathcal{I}^{\rm fin}=0$.) Second, in Appendix \ref{app:cf} we check that when using the finite expressions for the $C_{\rm F}$-terms  in Eqs.~\eqref{Ifin} and \eqref{Ivfin} to compute the integral over $\xi$ in \eqn{nlocf}, one does not generate any longitudinal logarithm. As shown by the above arguments and also by the manipulations in Appendix \ref{app:cf}, these  ``good'' properties are guaranteed to hold with either a fixed coupling, or a running-coupling $\abar(k_{\perp})$ with the scale set by the external transverse momentum. However, as we shall shortly argue, they would {\it not} hold anymore if the scale for the RC is rather chosen in {\it coordinate-space}. 

Let us first observe that a coordinate-space RC prescription would not be very natural in the present context, where the subtraction of the collinear divergences has been performed in momentum space. Indeed, the terms affected by this subtraction (the first term in \eqn{Ifin} and the whole term in \eqn{Ivfin}) cannot be written as a double integration in coordinate space --- unlike the original, singular, expressions for $\mathcal{I}$ and $\mathcal{I}_v$, cf.  \eqn{Ico}. As a consequence, we have no control anymore on the sizes of the ``daughter'' dipoles (the coordinate $\bx$ in  \eqn{Ico}). Thus, if one insists in using a running coupling with a transverse coordinate scale, then the only option is $\abar(r_{\perp})$. By inspection of Eqs.~\eqref{Ifin} and \eqref{Ivfin}, it is clear that such a choice would entail severe ambiguities. For instance, in one inserts the coupling $\abar(r_{\perp})$ inside the integral over $\br$ in the second equality of \eqn{Ivfin}, then one violates the equivalence with the expression shown in the first equality there (the {\it a priori} result of the collinear subtraction in momentum space). We are facing here exactly the same ambiguity as previously discussed in relation with Eqs.~\eqref{nk}--\eqref{nr}.  From that example, we know that the use of the RC $\abar(r_{\perp})$ would introduce the problem of the ``fake potential''. This problem is illustrated in Fig.~\ref{fig:i-ratios-d}.B, which shows the ratio $\mathcal{I}_v^{pos}/\mathcal{I}_v^{mom}$ between the results for $\mathcal{I}_v$ obtained with the position-space RC $\abar(r_{\perp})$ and the momentum-space RC $\abar(k_{\perp})$; one can see that $\mathcal{I}_v^{pos}$ has a different sign and also a different power tail at high $k_\perp$ as compared to $\mathcal{I}_v^{mom}$ --- in agreement with the discussion around Eqs.~\eqref{aft}--\eqref{nrres}.

An equally severe ambiguity, leading again to unphysical results, occurs if one tries to use the position-space RC $\abar(r_{\perp})$ in \eqn{Ifin} for the ``real'' term $\mathcal{I}^{\rm fin}$. The last term there is exactly the same as the third term in \eqn{Ico}, hence it can still be written as a double integral over $\br$ and $\bx$. So, it is in principle possible to use the RC $\abar(r_{\perp})$ in all the terms appearing in \eqn{Ifin} (inside the respective integrations, of course). However, if one does so, one obtains the results shown in Fig.~\ref{fig:i-ratios-d}.A, which differ by several orders of magnitude (and also in sign) from those obtained with the momentum-space RC $\abar(k_{\perp})$. Besides, the use of the position-space RC $\abar(r_{\perp})$ would spoil the cancellation between the two terms in \eqn{Ifin} in the limit where there is no scattering. It would furthermore spoil the cancellation between $\mathcal{I}^{\rm fin}$ and $\mathcal{I}_v^{\rm fin}$ in the limit $\xi\to 1$ (thus generating a spurious longitudinal logarithm in the $C_{\rm F}$ terms); see the discussion in Appendix \ref{app:cf}.

\begin{figure*}[t]
\begin{center}
\includegraphics[width=0.55\textwidth,angle=0]{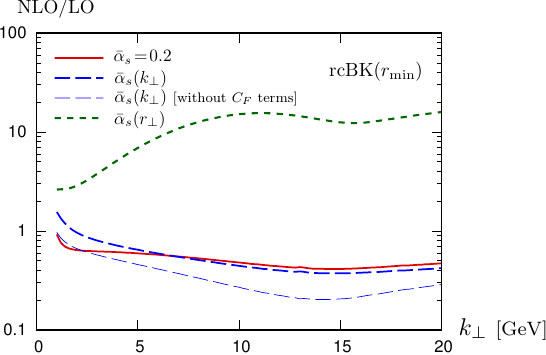}
\end{center}
\caption{\label{fig:NLOCF} The NLO quark multiplicity including both the $N_c$ terms and the $C_{\rm F}$ terms, normalised to the respective LO prediction and for three RC prescriptions, as indicated in the legend of the figure.  For comparison, we also show the corresponding result with the $N_c$ terms alone (the same as the curve ``$\abar(k_\perp)$'' in Fig.~\ref{fig:NLOratio}.A).
The evolution of the dipole $S$-matrix is always computed according to the rcBK equation with smallest-dipole prescription  $\abar(r_{\rm min})$.}
\end{figure*}

In Fig.~\ref{fig:NLOCF} we show the complete result for the NLO quark multiplicity as emerging from the present calculations, including both the $N_c$ terms and the $C_{\rm F}$ terms, and for three RC prescriptions: fixed coupling $\abar=0.2$, momentum-space RC  $\abar(k_{\perp})$, and the (pathological) coordinate-space RC $\abar(r_{\perp})$. To better appreciate the importance of the $C_{\rm F}$ terms, we also display the NLO result including just the $N_c$ terms (with momentum-space RC), that is, the curve that was denoted ``$\abar(k_\perp)$'' in Fig.~\ref{fig:NLOratio}.A. One thus sees that the effect of adding the $C_{\rm F}$ terms is indeed sizeable: the full NLO result is still smaller than the LO one, but the difference between the two is considerably reduced. A similar phenomenon --- namely the existence of  large 
$C_{\rm F}$  and $N_c$--corrections, which have opposite signs and largely cancel each other --- has been also noticed in the recent numerical calculation of DIS at NLO in Ref.~\cite{Ducloue:2017ftk}. That calculation too was based on a factorization scheme non-local in rapidity (similar to the one that we employed in this paper) together with the NLO impact factor for the virtual photon from  Refs.~\cite{Beuf:2016wdz,Beuf:2017bpd}.

To summarise, the problem with a coordinate-space prescription in the RC looks even more severe for the $C_{\rm F}$ terms than for the $N_c$ terms discussed previously. Unlike for the latter, we have no control on the size of the daughter dipoles anymore, so one cannot use the coupling $\abar(x_\perp)$ to avoid spurious contributions generated by the Fourier transform of the running coupling. The only reasonable choice seems to be a coupling running with a  transverse momentum scale, such as the momentum $k_\perp$ of the produced particle.

\section{\label{sec:conc} Conclusion and outlook}

In this paper we have continued our work on the NLO corrections to the single inclusive particle production at forward rapidities in high energy proton-nucleus collisions. For simplicity, we considered only the production of a quark, but ignored the gluon channel, as well as the subsequent fragmentation of the quark into hadrons. Our main focus was on the
choice of the scale for the running of the QCD coupling. This coupling appears in two distinct places in the calculation: the high-energy evolution of the dipole $S$-matrix, as described by the (NLO) BK equation, and the NLO impact factor.

The BK equation is generally solved by using the transverse {\it coordinate} space representation, whereas the NLO calculation of the impact factor is most naturally performed in the transverse {\it momentum} space. This naturally leads to the use of different running coupling prescriptions --- a coordinate-space prescription for the BK equation, but a momentum-space one for the NLO impact factor ---, which in turn introduces some ambiguities in practice (because the running of the coupling does not ``commute'' with the Fourier transform). It is then legitimate to ask whether one can eliminate such ambiguities by using a {\it unified} RC prescription throughout the calculation. A first proposal in that sense \cite{Ducloue:2017mpb}, using a coordinate-space formulation for the RC, was found to lead to unphysical results: the NLO correction to the quark multiplicity was larger by up to two orders of magnitude than the LO term and it also had the wrong sign compared to the calculation using the momentum-space RC prescription.

Here we have been able to understand the origin of the problem with the coordinate-space prescription proposed in  \cite{Ducloue:2017mpb}. We first observed that in the kinematical regime of interest, namely for relatively large transverse momenta $k_{\perp}\gtrsim Q_s$ for the produced quark, that prescription was essentially the same as the parent-dipole prescription  $\abar(r_{\perp})$, with $r_\perp\sim 1/k_\perp$ the transverse size of the parent dipole size. With this prescription, the RC $\abar(r_{\perp})$ acts as a spurious potential and leads to a large, but unphysical, contribution to the cross-section from primary gluons (daughter dipoles) with large transverse sizes $x_{\perp} \gg r_{\perp}$. This contribution is unphysical in that it violates transverse momentum conservation (i.e.~ the fact that the momentum of the primary gluon should match the one of the hard outgoing quark). In a correct calculation, this contribution would cancel due to the final Fourier transform, but this cancellation is badly spoiled by the fake ``potential'' $\abar(r_{\perp})$, which can transmit a large momentum. Regarding the $C_{\rm F}$-terms, the choice $\abar(r_{\perp})$ leads to additional problems: first, it generates non-zero, and thus unphysical, contributions in the limit of no scattering; second, it spoils the cancellation of the longitudinal logarithms (which in this factorization scheme should exist only in the $N_c$-terms).

We have also found that there is a meaningful scale choice in coordinate space, at least so long as the $N_c$-terms are considered. This is the ``daughter'' dipole prescription  $\abar(x_{\perp})$, which keeps intact the required cancellations. Numerically, we have seen that this choice gives almost the same results as the choice $\abar(k_{\perp})$, with the small difference being an acceptable scheme dependence. However, this choice $\abar(x_{\perp})$ is not possible in the $C_{\rm F}$-terms, where the coordinate $x_{\perp}$ of the primary gluon has been integrated over prior to the subtraction of the collinear divergencies (and hence this is not present anymore in the finite results). Thus, for these terms the use of the momentum space coupling $\abar(k_{\perp})$ seems to be mandatory. With the above choices, that is with $\abar(k_{\perp})$ or $\abar(x_{\perp})$ for the $N_c$-terms and $\abar(k_{\perp})$ for the $C_{\rm F}$-terms, we have numerically computed the quark multiplicity at NLO. We have found that the NLO correction reduces the LO result by 25-30\%, as one expects in a controlled pQCD calculation and in this regime of transverse momenta. Also we have found that the NLO $N_c$ and $C_{\rm F}$ contributions are of the same order, as also observed in the NLO calculation of the structure functions in DIS.

\begin{acknowledgments}

This work used computing resources from CSC -- IT Center for Science in Espoo, Finland.
A.H.M. and D.N.T. would like to acknowledge l'Institut de Physique Th\'eorique de Saclay for 
hospitality during the early stages of this work. This work of T.L. is supported by the Academy of Finland, projects 273464 and 303756. The work of T.L. and B.D. has been supported in part by the European Research Council, grant ERC-2015-CoG-681707. The work of B.D., E.I. and G.S. 
 is supported in part by the Agence Nationale de la Recherche project 
 ANR-16-CE31-0019-01.   The work of A.H.M.
is supported in part by the U.S. Department of Energy Grant \# DE-FG02-92ER40699.

\end{acknowledgments}

\appendix

\section{On the running coupling scale in the BK equation}
\label{app:bal}

Here we shall give a very simple argument leading to the appropriate choice of the running coupling scale in the BK equation. We recall that for a fixed coupling the latter reads
\begin{align}
\label{app-bkco}
 S(\br,X_g) = S(\br,X_0) + 
 \abar \int\limits_{X_g}^{X_0} 
 \frac{\dif X}{X}
 \int \frac{\dif^2 \bx}{2\pi}\,
 \frac{\br^2}{\bx^2(\br+\bx)^2}
 \left[ S(-\bx,X) S(\br+\bx,X) - S(\br,X)\right].
 \end{align}
The $\bx$-integration in \eqn{app-bkco} is convergent and the generic length scale for the coupling seems to be the ``external'' dipole size $r_{\perp}$. However, this is not always correct and to see that let us first decompose the dipole kernel (multiplied by the running coupling) as follows
 \beq
 \label{dipdec}
 \abar\, \frac{\br^2}{\bx^2(\br+\bx)^2}=
 \abar\, \frac{1}{\bx^2} + \abar \, \frac{1}{(\bx+\br)^2} - 
 \abar\,\frac{2 \bx \cdot (\br+\bx)}{\bx^2 (\br+\bx)^2}.
 \eeq
In the above consider the first term which arises from diagrams in which the soft gluon with transverse position $\bx$ has been emitted and reabsorbed by the same fermion of the parent dipole, the one located at $\bm{0}$. When $x_{\perp} \ll r_{\perp}$, the momentum flowing through the corresponding loop is $k_{\perp}\sim 1/x_{\perp}$ (it doesn't ``know'' anything about the other fermion located at $\br$). In that case, it is natural to choose $x_{\perp}$ as the proper length scale, that is 
 \beq
 \label{alphax}
 \abar\, \frac{1}{\bx^2}
 \to
 \abar(x_{\perp})\, \frac{1}{\bx^2}
 \quad {\rm for} \;\; x_{\perp} \ll r_{\perp}.
 \eeq
Accordingly, the coupling should be moved inside the integrand in \eqn{app-bkco}. Similarly, we find that the coupling in the second term in \eqn{dipdec} should be $\abar(|\br+\bx|)$ in the regime $|\br+\bx| \ll r_{\perp}$. Regarding the last term in \eqn{dipdec}, the argument cannot be applied since it stands for diagrams in which the soft gluon is emitted and absorbed by different fermions of the dipole. Furthermore, such a term is less singular when $\bx \to \bm{0}$ or $\br+\bx \to \bm{0}$, and thus its contribution in these regimes is strongly suppressed. Thus, for this third term one naturally keeps the coupling $\abar(r_{\perp})$.

To be more specific, let us give a concrete example which shows that (only) the above choices lead to the correct pQCD results. We consider that the target off which the dipole $\br$ scatters  is itself a dipole, with much smaller size $\br_0$ (i.e. $r_{0\perp} \ll r_{\perp}$). Then the BK equation reduces to the BFKL one, and moreover the dominant parts in the $\bx$-integration are two equal contributions from the \emph{collinear} regimes $r_{0\perp} \ll x_{\perp} \ll r_{\perp}$ and  $r_{0\perp} \ll |\br+\bx| \ll r_{\perp}$. At tree level the scattering amplitude $T=1-S$ reads $T(\bx,\br_0,X_0) \sim A r_{0\perp}^2$, and the first iteration gives 
 \begin{align}
 \label{tdla}
 T(\br,\br_0,X_g) = A r_{0\perp}^2 +
 A r_{0\perp}^2 \int\limits_{X_g}^{X_0} \frac{\dif X}{X}
 \int\limits_{r_{0\perp}^2}^{r_{\perp}^2} 
 \frac{\dif x_{\perp}^2}{x_{\perp}^2}\,   \frac{1}{\bar{b}
 \ln ({1}/{x_{\perp}^2\Lambda^2})}
 =A r_{0\perp}^2 
 \left[1 +  \frac{1}{\bar{b}} \ln \frac{\ln(1/r_{0\perp}^2\Lambda^2)}{\ln(1/r_{\perp}^2\Lambda^2)} \ln\frac{X_0}{X_g}\right].  	
 \end{align}
Clearly, the above result exponentiates and leads to the well-known double logarithmic series with running coupling. Similarly to \eqn{tdla} which is valid when $r_{0\perp} \ll r_{\perp}$, we can derive an analogous expression when  $r_{\perp} \ll r_{0\perp}$ and one can easily be convinced that the correct choice becomes $\abar(r_{\perp})$ as argued in the beginning of this appendix. 

Thus, it only remains to match the $\abar(x_{\perp})$ behavior for $x_{\perp} \ll r_{\perp}$ given in \eqn{alphax}, with the proper $\abar(r_{\perp})$ behavior when $x_{\perp} \gg r_{\perp}$. Since the way to do this matching is not unique, we must choose a \emph{prescription} and in the following we discuss two of them.

\noindent \emph{(i) The Balitsky prescription}: One way to extend the validity of \eqn{alphax} for any $\bx$ in a continuous way is
 \beq
 \label{abalx2}
 \abar\, \frac{1}{\bx^2}
 \to
 \frac{1}{\bx^2}\,
 \frac{\abar(x_{\perp}) \abar(r_{\perp})}{\abar(|\br+\bx|)},
 \eeq
and similarly for the second term in \eqn{dipdec}. For the third term we just keep  $\abar(r_{\perp})$ as already explained, while we furthermore rewrite the inner product  as $2 \bx \cdot(\br + \bx) = (\br+\bx)^2 +\bx^2 - \br^2$. Then we find that the kernel of the BK equation becomes
\beq
 \label{abal}
 \abar\, \frac{\br^2}{\bx^2(\br+\bx)^2}
 \to
 \abar(r_{\perp})
 \left\{
 \frac{\br^2}{\bx^2(\br+\bx)^2}
 + \frac{1}{\bx^2}
 \left[\frac{\abar(x_{\perp})}{\abar(|\br+\bx|)} -1 \right]
 +\frac{1}{(\br+\bx)^2}
 \left[\frac{\abar(|\br+\bx|)}{\abar(x_{\perp})} -1 \right]
 \right\},
 \eeq   
which is the well-known Balitsky prescription \cite{Balitsky:2006wa}. 

\noindent \emph{(ii) The minimum dipole prescription}: Another simple way to extend \eqn{alphax} reads
 \beq
 \label{minx2}
 \abar\, \frac{1}{\bx^2}
 \to
 \abar(r_{\rm min})\,
 \frac{1}{\bx^2}
 \eeq
and similarly for the second term in \eqn{dipdec}, where $r_{\rm min} = {\rm min}\{r_{\perp},x_{\perp},|\br+\bx|\}$. We are allowed to do the same replacement also for the last term in \eqn{dipdec}, since its contribution in the regimes $x_{\perp} \ll r_{\perp}$ and $|\br+\bx| \ll r_{\perp}$ is anyway suppressed. Thus we see that the kernel of the BK equation simply becomes
\beq
 \label{amin}
 \abar\, \frac{\br^2}{\bx^2(\br+\bx)^2}
 \to
 \abar(r_{\rm min})\, \frac{\br^2}{\bx^2(\br+\bx)^2},
\eeq
which is naturally called the minimum dipole prescription (see for example \cite{Iancu:2015joa}).

\section{On the properties of the $C_{\rm F}$ terms}
\label{app:cf}

Here we would like to check that the NLO contribution of the  $C_{\rm F}$ terms, as given by the integral over $\xi$ in \eqn{nlocf}, when used with the finite expressions for $\mathcal{I}^{\rm fin}$ and $\mathcal{I}_v^{\rm fin}$ shown in  Eqs.~\eqref{Ifin}--\eqref{Ivfin}, does indeed not generate any (leading) large longitudinal logarithms. As already said, this property is trivially satisfied before the DGLAP evolution is subtracted, cf.~Eqs.~\eqref{I} and \eqref{Iv}. Furthermore, it is even valid after the subtraction of the DGLAP evolution when plus distributions are used (cf.~Eq.~(43) in \cite{Chirilli:2011km,Chirilli:2012jd} or Eq.~(14) in \cite{Ducloue:2016shw}). However, it is not so obvious that this is true also in the current framework.

To this end we shall show that the difference $\delta\mathcal{I}^{\rm fin} \equiv \mathcal{I}^{\rm fin}-\mathcal{I}_v^{\rm fin}$ vanishes in the limit $\xi\to 1$. Notice that, when taken separately, both $\mathcal{I}^{\rm fin}$ and $\mathcal{I}_v^{\rm fin}$ are (logarithmically) divergent in that limit. This is manifest in \eqn{Ivfin}, which includes a term proportional to  $\ln(1-\xi)^2$, but it is
also true for the second term in \eqn{Ifin}, as it can be easily seen by replacing $\xi\to 1$ under the integration there. Thus in what follows we would like to show not only that these divergences mutually cancel, but also that the finite reminder vanishes as $\xi\to 0$.

For simplicity, we shall set $\xi=1$ in all the terms which are manifestly finite in this limit. For instance, instead of  \eqn{Ifin}, we shall consider
\beq
\label{Ifin1}
	\mathcal{I}^{\rm fin}(\xi) \to
	\abar \int \frac{\dif^2\br}{2\pi}\,
	S(\br)\ln \frac{c_0^2}{\br^2 \mu^2}
	\rme^{-\rmi \bk \cdot \br} 
	- 2 \abar \int \frac{\dif^2 \bq}{(2\pi)^2}
	\frac{(\bk - \xi \bq)\cdot(\bk-\bq)}{(\bk - \xi \bq)^2(\bk-\bq)^2}
	\mathcal{S}(\bq),
\eeq
which indeed captures the dominant behavior of $\mathcal{I}^{\rm fin}(\xi)$ near $\xi=1$.
(For the purposes of this Appendix, we will suppress some of the arguments in the quantities under consideration.) Concerning $\mathcal{I}_v^{\rm fin}$, we use Eq.~(30) of \cite{Chirilli:2011km,Chirilli:2012jd} to express the $\ln(1-\xi)^2$ in terms of a momentum integration, thus getting
\beq	
	\label{Ivfin1}
    \mathcal{I}_v^{\rm fin}(\xi) =
	\abar \ln \frac{\bk^2}{\mu^2} \int \frac{\dif^2\br}{2\pi}\,
	S(\br)
	\rme^{-\rmi \bk \cdot \br}
	-2 \abar \int \frac{\dif^2 \bq}{(2\pi)^2}\,
	\mathcal{S}(\bk)
	\left[ 
	\frac{(\bk-\bq)\cdot (\xi \bk-\bq)}{(\bk-\bq)^2(\xi \bk-\bq)^2} + \frac{\bq \cdot (\bk-\bq)}{\bq^2(\bk-\bq)^2}
	\right].
\eeq
It is now easy to check that the divergences at $\xi=1$ developed by the above integrals over $\bq$ cancel indeed between \eqref{Ifin1} and \eqref{Ivfin1}, as anticipated. It remains to show that the reminder approaches to zero when $\xi\to 1$. Combining Eqs.~\eqref{Ifin1} and \eqref{Ivfin1} and letting $\xi\to 1$, we deduce
\beq
 \label{dI1}
 \delta\mathcal{I}^{\rm fin}(1) = 
 \abar \int \frac{\dif^2\br}{2\pi}\,
	S(\br)\ln \frac{c_0^2}{\bk^2 \br^2}
	\rme^{-\rmi \bk \cdot \br}
 - \abar \int \frac{\dif^2 \bq}{(2\pi)^2}\,
 \frac{1}{(\bk-\bq)^2} \left[ 
 \mathcal{S}(\bq) - \frac{\bq\cdot \bk}{\bq^2} \mathcal{S}(\bk)
 \right].	 
\eeq
 Writing $\mathcal{S}(\bq)$ and $\mathcal{S}(\bk)$ as inverse Fourier transforms (after also making a shift $\bq \to \bq + \bk$), the second term in \eqn{dI1} becomes
 \beq
 \label{dI1b}
 \delta\mathcal{I}^{\rm fin}(1)|_{\rm 2nd} = 
 -\abar \int \frac{\dif^2\br}{2\pi}\,
	S(\br) \rme^{-\rmi \bk \cdot \br}
 \left[\frac{1}{\pi} \int \frac{\dif^2 \bq}{\bq^2}\, 
 \rme^{-\rmi \bq \cdot \br} 
 -\frac{1}{2\pi}
 \int \dif^2 \bq\, \frac{\bk^2}{\bq^2 (\bk-\bq)^2}
 \right],
 \eeq
where we simplified the last term by using $-2 \bq \cdot \bk = (\bk-\bq)^2 - \bq^2 -\bk^2$. The momentum space integrals in \eqn{dI1b} are individually divergent, but their sum will be finite. They can be performed using dimensional regularisation (the exact same integrals are found in \cite{Chirilli:2011km,Chirilli:2012jd}) and give
 \begin{align}
 \label{dimreg}
 &\frac{\mu^{2\epsilon}}{\pi}
 \int \frac{\dif^{2-2\epsilon} \bq}{\bq^2}\,\rme^{-\rmi \bq \cdot \br}
 = -\frac{1}{\epsilon} - \gamma_{\ssE} - \ln \frac{\mu^2 \br^2}{4\pi} + \mathcal{O}(\epsilon),
 \nn
 &\frac{\mu^{2\epsilon}}{2\pi}
 \int \dif^{2-2\epsilon}\bq \, 
 \frac{\bk^2}{\bq^2 (\bk-\bq)^2}
  = -\frac{1}{\epsilon} + \gamma_{\ssE} + \ln \frac{\pi \bk^2}{\mu^2} + \mathcal{O}(\epsilon),
 \end{align}
so that \eqn{dI1b} finally becomes
 \beq
 \label{dI1bfin}
 \delta\mathcal{I}^{\rm fin}(1)|_{\rm 2nd} = 
 -\abar \int \frac{\dif^2\br}{2\pi}\,
	S(\br) 
	\ln\frac{4 \rme^{-\gamma_{\ssE}}}{\bk^2 \br^2}
	\rme^{-\rmi \bk \cdot \br}.
 \eeq
This exactly cancels the first term in \eqn{dI1}, as anticipated; to conclude,
 \beq
 \label{dI10}
 \delta\mathcal{I}^{\rm fin}(1) =0.
 \eeq
Now, since we are interested in the $\xi$-dependence of the integrand in the $C_{\rm F}$ terms, let us write \eqn{nlocf} in the compact form
 \beq
 \label{nlocfcom}
 \frac{\dif N^{C_{\rm F}}}{\dif^2\bk\, \dif \eta} =
 \int \dif \xi \,\frac{1+ \xi^2}{1-\xi}\, f(\xi)\, 
 \mathcal{I}^{\rm fin}(\xi)
 - \int \dif \xi \,\frac{1+ \xi^2}{1-\xi}\, f(1)\, 
 \mathcal{I}_v^{\rm fin}(\xi).
 \eeq
We shall assume that the function $f(\xi)$ (i.e.~the quark pdf) close to $\xi=1$ can be expanded as
 \beq
 f(\xi) = f(1) - (1-\xi) f'(1) + \dots,
 \eeq
so that the quark multiplicity can be approximated by
 \beq
 \label{nlocfcomfin}
 \frac{\dif N^{C_{\rm F}}}{\dif^2\bk\, \dif \eta} \simeq
 \int \dif \xi \,(1+ \xi^2)f(1) \,  
 \lim_{\xi \to 1}\frac{\mathcal{I}^{\rm fin}(\xi) - \mathcal{I}_v^{\rm fin}(\xi)}{1-\xi}
 - \int \dif \xi \,(1+ \xi^2) f'(1)\, 
 \mathcal{I}^{\rm fin}(\xi).
 \eeq
Given our result in \eqn{dI10} and the analytic structure of $\mathcal{I}^{\rm fin}(\xi)$ and $\mathcal{I}_v^{\rm fin}(\xi)$, it is reasonable to assume that their difference vanishes linearly with $1-\xi$ as $\xi$ approaches the unity. Moreover, $\mathcal{I}^{\rm fin}(\xi)$ (like $\mathcal{I}_v^{\rm fin}(\xi)$) diverges only logarithmically as $\xi \to 1$, thus such a divergence is clearly innocuous when integrated over $\xi$. Hence, the $\xi$-integration in the $C_{\rm F}$ terms in \eqn{nlocfcomfin} does not give rise to a large longitudinal logarithm.  

\bibliographystyle{utcaps}
\providecommand{\href}[2]{#2}\begingroup\raggedright\endgroup

\end{document}